\documentclass[useAMS,usenatbib,usegraphicx]{mn2e}
\usepackage{epsfig}

\DeclareRobustCommand{\ion}[2]{%
\relax\ifmmode
\ifx\testbx\f@series
{\mathbf{#1\,\mathsc{#2}}}\else
{\mathrm{#1\,\mathsc{#2}}}\fi
\else\textup{#1\,{\mdseries\textsc{#2}}}%
\fi}

\newcommand{\vsini}{$v \sin i$}
\usepackage[usenames]{color}
\newcommand{\changea}{} 
\newcommand{\changec}{}

\def\teff{\hbox{$\,T_{\rm eff}$}}
\def\kms{\hbox{$\,{\rm km}\,{\rm s}^{-1}$}}
\def\ms{\hbox{$\,{\rm m}\,{\rm s}^{-1}$}}

\def\cd{\hbox{$\;{\rm c}\,{\rm d}^{-1}$}}

\def\degr{\hbox{$^\circ$}}
\def\logg{\hbox{$\log$\,g}}
\def\halpha{\hbox{${\rm H}\alpha$}}

\def\bet{\hbox{$\beta$}}

\def\arcmin{\hbox{$^\prime$}}
\def\arcsec{\hbox{$^{\prime\prime}$}}
\def\sun{\hbox{$\odot$}}

\def\teff{\hbox{$\,T_{\rm eff}$}}

\def\kms{\hbox{$\,{\rm km}\,{\rm s}^{-1}$}}

\title[17 Ap stars with magnetically resolved lines]{Discovery of 17 new sharp-lined
Ap stars with magnetically resolved lines\thanks{Based on observations 
collected at the  European Southern Observatory, Paranal, Chile, as part of 
programmes 078.D-0080(A), 078.D-0192(A), 072.D-0138(A).}}
\author[L. M. Freyhammer et al.]
 {L. M. Freyhammer$^{1}$\thanks{E-mail: lmfreyhammer\,@\,uclan.ac.uk},
 V. G. Elkin$^{1}$, D. W. Kurtz$^{1}$, G. Mathys$^{2}$ and P. Martinez$^{3}$
 \newauthor{}\\
 $^{1}$Centre for Astrophysics, University of Central Lancashire, Preston
 PR1 2HE\\
 $^{2}$European Southern Observatory, Casilla 19001, Santiago 19, Chile\\
 $^{3}$South African Astronomical Observatory (SAAO), PO Box 9,
Observatory 7935, South Africa}

\begin{document}

\date{Draft \today ;  Accepted . Received ; in original form }

\pagerange{\pageref{firstpage}--\pageref{lastpage}} \pubyear{2008}

\maketitle

\label{firstpage}

\begin{abstract}
Chemically peculiar A stars (Ap) are extreme examples of the interaction of
atomic element diffusion processes with magnetic fields in stellar 
atmospheres. The rapidly oscillating Ap stars provide 
a means for studying these processes in 3D and are at the same time
important for studying the pulsation excitation mechanism in A stars.
As part of the first comprehensive, uniform, high resolution 
spectroscopic survey of Ap stars, 
which we are conducting in the
southern hemisphere with the Michigan Spectral Catalogues as the basis of target 
selection, we report here the discovery of 17 new magnetic
Ap stars having spectroscopically resolved Zeeman components from which 
we derive magnetic field moduli in the range $3-30$\,kG. Among these
are 1) the current second-strongest known magnetic
A star, 2) a double-lined Ap binary with a magnetic component
and 3) an A star with particularly peculiar and variable abundances. Polarimetry
of these stars is needed to constrain their field geometries and 
to determine their
rotation periods. We have also obtained an additional measurement 
of the magnetic field of the Ap star HD\,92499.
\end{abstract}


\section{Introduction}
The study of chemically peculiar A stars with magnetic fields has 
ramifications for several other branches of astrophysics.  The 
interaction among strong magnetic fields, atomic diffusion and energy 
transfer in (or above) the upper atmospheres of non-degenerate stars has 
direct implications for observed stellar abundances and for the 
instability of stellar 
pulsations in $\beta$\,Cephei and sdB stars. The magnetic field lines guide 
the diffusing elements, so that some ions will be 
concentrated where the field lines 
are vertical while others will group where the lines are horizontal (e.g. 
\citealt{michaud81}).  The very different magnetic field strengths, 
orientations and 
geometries, as well as the great variety of abundance distributions 
for Ap stars are therefore particularly informative.

Magnetic fields affect spectroscopic lines
through, e.g., broadening, intensification and Zeeman splitting of the
intrinsic lines \citep{mathys89}.
Several cool Ap stars show light and spectral variability that follow the stellar
rotation, and such stars are known as
 $\alpha^2$ Canum Venaticorum ($\alpha^2$\,CVn) variables.
\citet{wolfetal71} demonstrated that the observed light, spectrum and magnetic 
variations are related. In particular, they suggested that photometric variability
may result from a redistribution of flux either by variations in line blanketing
or opacity, such as enhanced absorption at rare earth maximum. 
The light variability may vary with that of the rare earths which appear to be
concentrated predominantly near the region of the strongest magnetic field 
or that of the negative pole. Thus can light variability be linked to magnetic field 
variability which, for oblique rotators, follows the stellar rotation.

Magnetic activity cycles similar to the solar 11-y cycle are 
observed in F--M stars \citep{baliunas95} and are related to chromospheric 
activity. However, chromospheres have not been unambiguously detected in
Ap stars. Finally, magnetic fields drive winds through Alfv\'en waves  
or other magneto-hydrodynamic  waves \citep{linsky04}.

The chemically peculiar B, A and F stars 
(henceforth Ap stars) have globally organised, typically dipole or 
quadrupole fields that with an axis obliquely inclined 
to the rotation axis of the star so that, for dipolar fields,  one or both 
magnetic poles come into, or out of, view with rotation. The full 
geometry of the field may hence be visible for stars with favourably 
oriented rotation and magnetic axes, which in turn makes it possible 
to use the 
Zeeman Doppler imaging technique to create magnetic field maps. Among 
the Ap stars is Babcock's star = HD\,215441 \citep{babcock60}, the 
non-degenerate star with the strongest field known, 34.4\,kG. The only 
other published cases with extremely strong fields are:
HD\,137509, $\langle B\rangle=29$\,kG \citep{kochukhov06}, 
HD\,154708, $\langle B\rangle= 24.5$\,kG \citep{hubrigetal05} and
HD\,178892, $\langle B\rangle=18.0$\,kG \citep{ryabchikova06}.

The atmospheres of Ap stars are complex to interpret. 
As a consequence of vertical abundance stratification, and 
non-standard temperature gradients, 
even theoretical modelling of hydrogen line profiles, 
such as the core-wing anomaly 
in the H$\alpha$ lines \citep{cowleyetal01}, is not yet completely 
successful \citep{kochukhovetal02}. However, 
the observational and theoretical efforts in understanding the Ap 
atmospheric processes may eventually prove `worth the candle'
thanks to their general applicability; the Ap stars are the
most extreme examples of magnetic fields and atomic diffusion
in non-degenerate stars, processes common in most other stars. 
Examples of studies of atmospheric abundance stratification in a
cool Ap star 
subgroup, the rapidly oscillating Ap (roAp) stars are given by 
\cite{wadeetal01,ryabchikovaetal02,ryabchikovaetal05}.
These studies are in agreement
with each other, in general: Fe is concentrated by gravitational settling
in the observable layer between $-1<\log\tau_{\rm {5000}}<0$ and Pr and
Nd are concentrated by radiative levitation above $\log\tau_{\rm {5000}}=-5$. 
The Pr and Nd forming layers are above the line forming layer
of the narrow core of the H$\alpha$ line which in standard A star models 
is in the range $-4<\log\tau_{\rm {5000}}<-2$.

Magnetic field measurements are mostly based on difficult, indirect 
measurements, such as 
magnetic broadening of spectral lines and spectropolarimetric observations 
of longitudinal magnetic fields or the line-of-sight field component.
 The mean longitudinal magnetic field (or, the longitudinal field) 
$\langle B_{\rm z}\rangle$ is a weighted average over the
visible stellar disk of the component of the magnetic vector along the line of sight
\citep{mathys89}, and is typically at least 3 times weaker than the
mean magnetic field modulus $\left<B\right>$ (see Sect.\,\ref{sec:mag} for
definition of this).
  The first large survey for magnetic fields in non-degenerate stars was done by 
 \citet{babcock58}. It has since been followed by many other studies
 of individual stars or groups of stars (e.g., 
 \citealt{borraetal79,bohlenderetal93}). Recent 
 searches for magnetic stars were carried out by 
 \citet{kudryavtsevetal06} who discovered $\sim 70$  
 magnetic stars, and by \citet{hubrigetal06} 
 who discovered 57 magnetic Ap stars.
 The total number of firmly established magnetic chemically peculiar
 main-sequence stars is currently $\sim 350$ (Romanyuk, 2008, in prep.)
  For a significant fraction of these magnetic stars the variation in
 the longitudinal magnetic field has been measured as a function of rotation 
 period (see, e.g., \citealt{mathys91}).

 Such curves provide valuable information about the global geometrical 
 structure of magnetic fields.
Better and more reliable measurements are necessary to obtain 
a complete understanding of the many phenomena related to magnetic fields.  
For this purpose, magnetic 
Ap stars with resolved magnetically split lines provide extremely favourable 
conditions, since one can determine in a straightforward, mostly 
approximation-free, model-independent manner, and with particularly good 
precision, the mean magnetic field modulus \citep{mathysetal97}.
However, in spite of many studies (e.g., \citealt{mathysetal97}; 
\citealt{hubrigetal07}),
only 51 such stars were known \citep{hubrigetal07}, prior to this work.
Our discovery of 17 new such stars therefore significantly increases the 
number known. 

At present, 40 roAp stars are known (see, e.g., 
\citealt{kurtzetal06b,gonzalezetal08}), although
several surveys have searched for rapid pulsation in Ap stars, such as
\citet{nelsonetal93, martinezetal94, handleretal99, ashokaetal00, weissetal00,
dorokhovaetal05}. For a (non-exhaustive) list of spectroscopic studies of 
roAp stars, see \citet{kurtzetal06a}.
The roAp stars are thought to be characterised by, e.g., the H$\alpha$\
{\em core-wing anomaly} and by an {\em ionization disequilibrium}
for Nd\,\textsc{ii} and Nd\,\textsc{iii} and for Pr\,\textsc{ii} and
Pr\,\textsc{iii} \citep{ryabchikovaetal04}, the latter caused
by stratification of those two elements to levels above $\log \tau_{{5000}}
\sim -5$ by atomic diffusion. Other typical spectral characteristics are 
strong magnetic fields, wing-nib anomaly in \ion{Ca}{ii}\,K 
\citep{cowleyetal06}, and strong abundances of rare earth elements and 
relatively slow rotation. It is therefore possible to identify promising roAp 
star candidates with a single, high signal-to-noise ratio, 
high-resolution spectrum before spending 
time on a large telescope with fast spectroscopy to search for rapid pulsations.

In 2006 we therefore began a systematic survey of cool Ap stars in the 
photometric `Cape cool Ap star catalogue' \citep{martinez93} to identify 
roAp candidates based generally on a single high-resolution spectrum
 of each star. The Cape
catalogue gives Str\"omgren $uvby$ and $\beta$ photometry for over 500 (almost all) 
of the SrCrEu subclass of cool Ap stars listed in the Michigan spectral
catalogues, volumes $1-4$ \citep{houk78,houk82,houketal75,houketal88}.
Perhaps surprisingly -- given the importance of the Ap 
stars to stellar astrophysics, and the long history of their study -- there is no 
large-scale, uniform spectroscopic survey such as the one we are now conducting. 
We expect the data set to be a rich source of discoveries, and to be the basis of 
uniform statistical analyses of many of the astrophysically interesting 
characteristics of the class. 

Our first observing season is finished and of 140 stars, 
we identified
dozens of stars with magnetically intensified, broadened or even
resolved lines. Of these, 17 are new detections with magnetically 
resolved or partially resolved lines, in particular for the Zeeman doublet 
\ion{Fe}{ii}\,6149.258\,\AA\
(Figs\,\ref{fig:fe6149a}--\ref{fig:fe6149b}). 
As pointed out by \citet{mathysetal97}, this line is particularly 
important as a diagnostic for a magnetic field as the splitting
provides a direct measure of the mean magnetic field, and
furthermore because iron is usually rather homogeneously
distributed over the surface of Ap stars. 

This is the discovery paper for the new Ap stars with magnetically 
resolved lines, and we provide magnetic field strengths, projected rotation 
velocities and selected relative abundance estimates. 
Of particular interest, we discovered a new star with 
an extremely large magnetic field; a highly peculiar magnetic star; 
and an uncommon magnetic Ap star in a relatively close binary system.
We also re-observed the recently discovered magnetic Ap star 
HD\,92499 \citep{hubrigetal07} to check for stability of its 
magnetic field strength, abundances and radial velocity. 

In the following sections, we describe the selection and observation of the
targets along with the data reduction in Sect.\,2. Then follows the data analyses, 
including estimation of physical parameters and the magnetic measurements
in Sect.\,3.  Finally we discuss the results in Sect.\,4.

\section{Observations and data reduction}

\begin{table}
 \begin{minipage}{81mm}
\caption{{\normalsize \label{tab:obslog}Observing log indicating target, 
 equatorial coordinates, instrument (UVES, `U' or FEROS, `F'), mid-exposure 
 heliocentric Julian Date (HJD), exposure time, number of  co-added spectra, 
 if any, and $S/N$. The $S/N$ ratio were measured in the 2-D spectra around
 5500\,\AA.  For co-added spectra, HJD gives the 
 mid-series time, and the total exposure time and combined $S/N$ are given.
}}
  \begin{tabular}{@{}l@{~}l@{~~}c@{~\,}c@{\,}c@{\,}c@{~}l@{}r@{}}
  \hline
Star        & $\alpha_{2000.0}$&$\delta_{2000.0}$  &Ins.&{HJD}
&$t_{\rm exp}$& $n$& $S/N$\\ 
HD          &(h:m:s)     & (\degr:\,\arcmin\,:\,\arcsec\,)& & ($-$2450000)&(s)& &                
\\ \hline
     33629  & 05:10:05.8 & --33:46:46 & F&4138.56813  & 800  &  & 198 \\
     42075  & 06:07:36.9 & --26:37:16 & F&4138.62481  & 800  &  & 195 \\
            &            &            & F&4141.57144  & 800  &  & 247 \\
            &            &            & U&4172.50223  & 450  &  & 273 \\
     44226  & 06:19:34.8 & --25:19:42 & F&4138.65922  &1100  &  & 170 \\
     46665  & 06:33:41.1 & --22:41:45 & F&4138.69450  &1200  &  & 170 \\
            &            &            & U&4173.48552  & 460  & 2& 185 \\  
     47009  & 06:35:48.6 & --13:44:50 & F&4138.70824  & 800  &  & 159 \\
     52847  & 07:01:46.3 & --23:06:20 & F&4138.72195  & 300  &  & 165 \\
            &            &            & U&4173.50404  & 300  &  & 311 \\
     55540  & 07:12:30.4 & --21:03:54 & F&4138.75030  &1100  &  & 165 \\
            &            &            & F&4139.68374  &1200  &  & 223 \\
            &            &            & F&4141.55994  &1100  &  & 231 \\
            &            &            & U&4173.49684  & 350  &  & 177 \\
     69013  & 08:14:29.0 & --15:46:32 & F&4140.60574  & 900  &  & 153 \\
     72316  & 08:30:58.5 & --33:38:04 & F&4140.62692  & 500  &  & 171 \\
     75049  & 08:45:33.1 & --50:43:58 & F&4141.66711  &1500  & 2& 265 \\ 
            &            &            & U&4171.50465  & 900  & 4& 285 \\  
     88701  & 10:13:00.2 & --37:30:12 & F&4140.69708  & 720  &  & 157 \\
     92499  & 10:40:08.6 & --43:04:50 & F&4138.80452  &600   &  & 178 \\
            &            &            & F&4139.66861  &900   &  & 223 \\
            &            &            & F&4141.73381  &900   &  & 238 \\
            &            &            & U&4172.49146  &630   & 4& 281 \\  
     96237  & 11:05:34.0 & --25:01:09 & F&4138.82842  &1100  &  & 195 \\
            &            &            & U&4173.51530  &300   &  & 152 \\
     110274 & 12:41:31.0 & --58:55:24 & F&4138.88351  & 900  &  & 169 \\
            &            &            & F&4141.82828  & 900  &  & 192 \\
            &            &            & U&4172.90849  & 500  &  & 212 \\
     117290 & 13:30:13.2 & --49:07:59 & F&4139.76263  & 900  &  & 190 \\
            &            &            & F&4141.86589  & 900  &  & 221 \\
            &            &            & U&4171.92209  & 600  & 2& 233 \\  
     121661 & 13:58:42.4 & --62:43:07 & F&4141.78480  & 350  &  & 156 \\
            &            &            & U&4171.91084  & 600  & 2& 311 \\  
     135728AB& 15:17:38.8& --31:27:32 & F&4139.86100  & 500  &  & 188 \\
            &            &            & U&4172.92442  & 300  &  & 257 \\
     143487 & 16:01:44.2 & --30:54:57 & F&4140.82232  & 900  &  & 153 \\
            &            &            & U&4173.91731  &960   &12& 308 \\
 \hline
\hline
\end{tabular}
\end{minipage}
\normalsize
\end{table}

The observations were collected using the Fibre-fed Extended Range Optical
Spectrograph (FEROS) with the 2.2-m telescope at La Silla 
during four nights starting 2007 February 06. Follow-up observations were 
made for selected stars about one month later with the VLT UV-Visual Echelle 
Spectrograph (UVES) during three nights starting 2007 March 11 (see 
the logbook of observations in Table\,\ref{tab:obslog}).   Both instruments 
are echelle spectrographs and cover the wavelength ranges 3530--9220\,\AA\ at 
the spectral resolution $R=\lambda/\Delta\lambda=48000$ (FEROS) and 
4970--7010\,\AA\  (with a 60-\AA\  wide gap near 6000\,\AA) at 
$R=110000$ (UVES).  The UVES instrument was used with an image slicer and an 
0.3\,arcsec slit, while FEROS was used with a stellar fibre (1.8\,arcsec 
aperture) while the second (sky) fibre was not used.

The spectra were reduced to 1-D with the respective pipelines provided
by ESO with the standard calibration data (bias, flatfield and Thorium-Argon 
wavelength reference spectra).  As indicated in the observing log 
(Table\,\ref{tab:obslog}) spectra of the same star were co-added if
barycentric velocity corrections were small compared to the bin sizes 
of the spectra (that was typically the case within 40\,min). Table\,\ref{tab:obslog}  
gives
\changea{
the signal-to-noise ratio ($S/N$ henceforth) of the spectra 
measured in the 2-D spectra around 5500\,\AA\ (for co-added 1-D spectra, the stated $S/N$
is for the composites).}
The spectra were normalized to unity continuum level in an iterative manner: 
major slopes were eliminated with a spline fit to a stacked spectrum of 
stars with relatively many continuum windows, then a second
but more detailed spline fit was made and applied to the spectra
on a star-by-star basis. Finally, a detailed spline fit was made by
identifying continuum regions through comparison with a synthetic
spectrum. This process eliminated flux undulations of lengths
similar to the individual echelle orders, but also broad
depressions from the strongest diffuse interstellar bands and 
possibly some of the broadest depressions typical of cool Ap stars.
\section{Data analysis and results}
\subsection{Light variability}
\label{sec:asas}
The photometric databases ASAS (All Sky Automatic Survey,
\citealt{pojmanski02}) 
and the {\it Hipparcos} catalogue \citep{hip} were used to search the
18 targets for long-period (days to years) photometric variability. 
The ASAS light curves are sparsely sampled, typically
one measurement per $1-3$ nights (causing obvious 1\,\cd\ aliases; \cd\ is
cycles per day), but 
cover $3-5$ years with $300-500$ measurements and are therefore well suited
to check for variability linked with stellar rotation. Of 5 available 
apertures we used the largest and for the $V$-band.
In seven cases (see notes in Table\,\ref{tab:targets}), periodic
variability was found above $4-5$\,mmag amplitude and interpreted
\changea{
as $\alpha^2$\,CVn rotational variability. 
All frequency searches}
were made using {\small PERIOD04} \citep{lenzetal05} up to
the pseudo-Nyquist frequency ($\sim 0.5$\,\cd\ for the ASAS data). 
The shortest
period found in these cases is a good candidate for the rotation period
for stars where only a single spotted region is within view. 
However, if
two such spots come into view during a rotation cycle (such
as orientations where both magnetic poles can be seen during
each rotation cycle), then the light curve becomes a double-wave (see
Fig.\,\ref{fig:HD88701}) which for similar spots may be
mistaken for a single sine wave, thus resulting in 
a mis-identification of only half the true rotation period.

\subsection{Line analysis}
For the purpose of spectral line identification, a synthetic comparison 
spectrum was produced with {\small SYNTH} \citep{piskunov92} using a Kurucz 
stellar atmosphere model.  Atomic line data were taken from the Vienna 
Atomic Line Database (VALD, \citealt{kupkaetal99}) for ions with 
increased abundances, 
mainly for Nd, Pr, Sr, Cr and Eu.  Other sources used for line data 
were the atomic database NIST\footnote{http://physics.nist.gov} and the  
Database on Rare Earth Elements at Mons University 
(DREAM\footnote{http://w3.umh.ac.be/$\sim$astro/dream.shtml}; 
\citealt{biemontetal99}) through its
implementation in the VALD. 
\begin{table*}
\caption{\label{tab:targets}
Main properties of the 18 magnetic Ap stars. The columns give: ($2-6$) 
Str\"omgren and $\beta$ 
indices from \citet{martinez93}; \changec{($7$) the revised {\it Hipparcos}
parallax $\pi$ \citep{vanleeuwen07}; ($8$) luminosity based on absolute visual
magnitude from \citet{gomezetal98} (except for HD\,88701 that
was derived directly, assuming no reddening), 
bolometric corrections based on the relation by 
\citet{landstreetetal07}, and $M_{{\rm bol},\sun}=4.72$.}
(9) Temperature derived from the $\beta$ index with the $c_0,\beta$ grid by 
\citet{moonetal85}; $\sigma(T_{\rm eff})=200$\,K, estimated error. (10) 
Projected rotation velocity \vsini; $\sigma$(\vsini)=1\,\kms, estimated error,
based on line-width measurements and synthetic, rotationally broadened line 
profiles compared to magnetically 
insensitive iron lines. \changea{(11) Mean magnetic field modulus from 
direct measurements of the $\lambda$6149\,\AA\ line, given as average
for all available spectra.
For HD\,96237, other lines were used instead (see Sect.\,\ref{sec:96237}).  
Error estimates are 
based on repeated measurements on different spectra (see 
Sect.~\ref{sec:individual} for more explanation).}  (12) Notes on individual 
stars.  $\alpha^2$CVn indicates $\alpha^2$ Canum Venaticorum variables, 
based on ASAS photometry, with minimum period indicated in parentheses.  
Photometric indices and parallax for the double-lined spectroscopic binary 
(`SB2') HD\,135728 are for the combined system.}

\begin{tabular}{@{}l@{~}c@{~~}c@{~~}c@{~~}c@{~~}c@{~~}c@{~~}c@{~~}c@{}c@{}r@{~~}l@ {}} \hline
Star   &$V  $&$b-y$  &$m_1$ &$c_1$&\bet &  $\pi$      &$\log L$&\teff&~\vsini& $\left<B\right>\phantom{xxx}$& Notes\\
HD     & mag & mag   & mag  & mag & mag & (mas)       &L$_\odot$& K & ~\kms& kG\phantom{xxxx}  &   \\  \hline
33629  &9.064& 0.200 &0.231 &0.683&2.784&             &         & 7570&  4   &$4.76\pm0.20$&   \\
42075  &8.968& 0.205 &0.260 &0.665&2.788&             &         & 7590&  1   &$8.54\pm0.02$&   \\
44226  &9.492& 0.161 &0.259 &0.771&2.842&             &         & 8060&  2   &$4.99\pm0.15$&   \\
46665  &9.441& 0.039 &0.255 &0.702&2.840&$2.11\pm1.16$&$1.50\pm0.24$&8050& 1 &$4.63\pm0.13$&  \\
47009  &9.070& 0.091 &0.201 &0.885&2.873&$2.75\pm1.25$&$1.74\pm0.31$&8280& 6 &$7.36\pm0.15$&  \\
52847  &8.157& 0.101 &0.334 &0.633&2.843&             &         & 8060&  1   &$4.44\pm0.01$&   \\
55540  &9.498& 0.026 &0.258 &0.774&2.868&             &         & 8230&  5   &$12.73\pm0.30$&$\alpha^2$CVn? \citep{hensbergeetal84}, \\ 
       &     &       &      &     &     &             &         &     &      &              & RV variable.\\ 
69013  &9.456& 0.296 &0.330 &0.400&2.772&             &         & 7470&  4   &$4.80\pm0.10$&   \\
72316  &8.804&--0.001&0.248 &0.836&2.840&$2.47\pm0.93$&$1.74\pm0.27$&8050&6  &$5.18\pm0.40$&   \\
75049  &9.090&--0.054&0.249 &0.914&2.906&             &         & 9700&  8   &$30.29\pm0.08$& $\alpha^2$CVn (5.28\,yr and 4.05\,d)\\ 
88701  &9.258& 0.002 &0.225 &0.830&2.846&$2.35\pm0.99$&$1.41\pm0.85$& 8080&7 &$4.38\pm0.35$& $\alpha^2$CVn (25.765\,d)\\
92499  &8.890& 0.179 &0.301 &0.615&2.812&$3.54\pm0.83$&$1.25\pm0.20$& 7810&2 &$8.20\pm0.13$& Magnetic$^a$. $\alpha^2$CVn ($>5$\,yr)\\
$96237_{\phi1}$&9.434&0.233&0.261&0.704&2.824&$1.53\pm1.15$&$1.61\pm0.29$&7930&5&2--3~~~~~~&$\alpha^2$CVn (20.9\,d)\\
$96237_{\phi2}$&     &     &     &     &     &             &             &7800&   &$2.87\pm0.30$&Rapidly changing, highly peculiar.\\
110274 &9.328& 0.195 &0.208 &0.805&2.853&$0.61\pm1.30$&$1.82\pm0.25$&8130& 1 &$4.02\pm0.38$& $\alpha^2$CVn ($265.3$\,d) \\
117290 &9.281& 0.152 &0.229 &0.879&2.867&             &         & 8230&  2   &$6.38\pm0.02$& $\alpha^2$CVn ($>5.7$\,yr)   \\
121661 &8.556& 0.036 &0.234 &0.805&2.866&             &         & 8230&  1   &$6.16\pm1.14$& $\alpha^2$CVn ($47.0$\,d). Magnetic$^b$. \\ 
135728A&8.602& 0.236 &0.203 &0.931&2.843&$4.27\pm1.17$&$1.58\pm0.25$&8060& 10& 0--2.5~~~~& SB2 binary.\\
135728B&     &       &      &     &     &             &         &     &  2   &$3.63\pm0.30$&     \\
143487 &9.420& 0.386 &0.262 &0.393&2.706&             &         & 6930&  2   &$4.23\pm0.07$&     \\
 \hline \hline
 \multicolumn{12}{l}{Measurements in literature: $a$) $\langle B\rangle=8.5\pm0.2$\,kG\,\citep{hubrigetal07}, 
  $b$) $\langle B_{\rm s}\rangle=2\,$kG and $\log g=4.6$\,\citep{northetal84}}   
\\ 
\end{tabular}
\end{table*}

\subsubsection{Temperatures and rotational velocities}
\label{sec:teffrot}

The spectra of Ap stars have flux distributions that are deformed by strong 
overabundances of some elements. 
Calibrations based on Str\"omgren indices may therefore 
not provide reliable estimates of temperatures and surface gravity \logg. 
The \bet\ index, however, remains largely unaffected by this, 
except in extreme cases -- HD\,101065, `Przybylski's star', being the 
most notorious example.  Effective 
temperatures were thus estimated with the $c_0,\beta$ grids of 
\citet{moonetal85} using \bet\ from Table\,\ref{tab:targets} (from 
\citealt{martinez93}). We did not use the value of $c_0$ in these grids,  
since $c_0$ is not a reliable indicator of luminosity in Ap stars because of the 
strong line blanketing in the Str\"omgren $v$ band; in any case, temperature in 
the relevant range is relatively insensitive to $c_0$.  We fixed 
\bet\ and assumed $4.0\le\log g\le4.5$. The resulting photometric 
temperatures are given in Table\,\ref{tab:targets}. These estimates depend 
on the actual evolutionary stage and contamination of the spectra by 
peculiar abundances, and stratification. The relative strengths of the 
hydrogen lines H8, H$\epsilon$ and of \ion{Ca}{ii}\,K are also temperature 
sensitive and indicate, for `normal' main-sequence stars, temperatures that 
are $500-1000$\,K higher for HD\,52847, HD\,55540, HD\,121661 HD\,143487, 
and $500-1000$\,K lower for HD\,42075, HD\,44226, HD\,92499, HD\,117290 and 
HD\,135728AB. 

\changec{
A few of the stars in Table\,\ref{tab:targets} have revised {\it Hipparcos}  
parallaxes \citep{vanleeuwen07} with relative trigonometric errors of $\sigma(\pi)/\pi=0.23$--0.75 
(2.13 for HD\,110274). Seven of these have distances and 
absolute magnitudes, $M_V$, given by \citet{gomezetal98}, while $M_V$ 
for HD\,88701 was calculated directly from the parallax assuming 
no reddening.  \citet{gomezetal98} used the statistical LM method 
\citep{lurietal96}, } which is
a maximum likelihood method that exploits a combination of proper motion 
and radial velocity data and trigonometric parallaxes
to obtain luminosity calibrations and improved distance estimates.
Combined with our photometric temperatures, these stars form a 
concentrated group in the ranges  $7810 \le T_{\rm  eff} \le 8280$\,K 
\changec{
and $0.1 \le M_V \le 1.5$, or $1.25 \le \log{L/{\rm L}_{\sun}} \le 1.82$ for
bolometric corrections from the relation by \citet{landstreetetal07} 
and using $M_{{\rm bol},\sun}=4.72$. 
Luminosity errors in  Table\,\ref{tab:targets} were estimated from distance errors
provided by \citet{gomezetal98}, 0.1\,mag error in the relation for bolometric correction, 
and temperature errors (through their influence on the calculated bolometric corrections).
Because of the small temperature range, the bolometric corrections 
are near zero and introduce errors of $\sim 0.1$ in $\log{L/{\rm L}_{\sun}}$.
 \citet{landstreetetal07} derived new bolometric corrections for Ap stars as a function of 
temperature, using temperatures based on photometric indices. 
They adopted uniform `somewhat optimistic' uncertainties of about 500\,K.
Photometric temperature
estimates are particularly problematic for Ap stars due to  their sensitivity to the large 
blanketing effects caused by strong absorption in (mostly) rare earth element lines.  
We find, however, a very good agreement between the predicted and 
measured bolometric correction for the roAp star $\alpha$\,Cir \citep{brunttetal08} with
\teff=7420\,K.
}
 
Keeping 
the considerable luminosity errors in mind, and that HD\,135728 is a binary, this suggests 
these are late main-sequence stars around the main-sequence turn-off. 
\changec{
Only a single luminous roAp star is known 
(see \citealt{elkinetal05} about the luminous
roAp HD\,116114 and \citealt{freyhammeretal08b} for a null-result for 
HD\,151878).
\citet{freyhammeretal08} excluded rapid oscillations 
for 9 luminous Ap stars to high precision, so it may be particularly difficult
to detect roAp stars among the sample here.
}

Projected rotation velocities (also in Table\,\ref{tab:targets}) were 
determined from the widths of the two \ion{Fe}{i} lines 
$\lambda\,$5434.52\,\AA\  and 6586.69\,\AA\ that are rather 
insensitive to magnetic 
broadening. A few rotational standards observed in the 2007 February run and 
some available UVES spectra of a few roAp stars with known rotation velocities 
were used as references for a FWHM--\vsini\ relation. These velocities were 
verified during the {\small SYNTHMAG} analysis (Sect.\,\ref{sec:abundance}), 
and found to agree within $\pm1\,\kms$.

\subsubsection{Abundance estimates}
\label{sec:abundance}
For the purpose of evaluating the level of chemical peculiarities and
ionization disequilibria, we made simple abundance estimates of Fe, Cr, Nd, Pr
and Eu. Synthetic spectra were produced with {\small SYNTHMAG} 
\citep{piskunov99} for $\log g = 4.0$ and 4.5, and the measured magnetic 
field strengths and photometric temperatures and then compared to the 
observed spectra for a range of Cr, Fe, Nd, Pr, Eu abundances.  
Because of the strong magnetic fields, we used model atmospheres 
from the online grid\footnote{Available at http:\//\//kurucz.harvard.edu}
 by \citet{castellietal03}
calculated with the {\small ATLAS9} code with  $\xi=2$\,\kms\ 
pseudo-microturbulence in order to simulate effects from magnetic 
intensification of spectral lines.  Model atmospheres with
increased solar metal abundance 
$\log{(N_{\rm{Z}}/N_{\rm Z,\sun})}=+0.5$ were used.  
For spectrum synthesis for the magnetic stars, we used 
$\xi=1$\,\kms\ as the {\small SYNTHMAG} code includes magnetic 
intensification by directly including magnetic field effects. This value
is slightly conservative, but is based on
on values used by other studies in the literature for magnetic Ap stars 
(such as 1--4\,\kms\ for 1--9\,kG fields: 
\citealt{ryabchikovaetal06b,ryabchikovaetal07,cowleyetal08}).

Atomic line data were 
taken from the VALD database using increased rare earth element 
abundances (identical for all models). The grid of models had 
$7000 \le T_{\rm  eff} \le 9500$\,K according to the photometric temperatures 
in Table\,\ref{tab:targets}, in steps of 500\,K. Abundances were computed
in steps of 0.1--0.5 dex but optimised on a case-by-case basis.
The simplified {\small SYNTHMAG} model of magnetic fields
is characterised by three components: radial field 
$B_{\rm r}$ (field component parallel to the line of sight), 
meridional $B_{\rm m}$ (field component parallel
to the surface at every surface point in the
plane-parallel atmosphere) and longitudinal
component $B_{\rm l}$ (assumed zero, as
is justified for the plane-parallel approximation).
For simplicity, magnetic fields were assumed to be 
combinations of equally strong radial and meridional fields based on the 
magnetic modulus values in Table\,\ref{tab:targets}. In a few cases, a 
longitudinal field was used instead of the meridional field.  Abundances of 
Fe, Cr, Nd, Pr and Eu were varied in ranges corresponding to those found by 
\citet{adelman73} in his study of 21 cool Ap stars.  
Instrumental broadening of 0.05\,\AA\ was adopted throughout, while 
$1-6$\,\kms\ macroturbulence was used to optimise the fitting of observed 
spectra with model spectra on a star-by-star basis. Surface 
gravity of $\logg=4.0$ was assumed, except for HD\,42075, HD\,55540, HD\,75049 
and  HD\,121661  where $\logg=4.5$ was used.  

Because of strong line blending, 
the common presence of unidentified lines in Ap spectra, and a typical 
$S/N \sim 200$ of our spectra, it is not always easy to identify continuum 
regions around spectral lines to be used in the abundance analysis. 
Therefore, rather than rely on equivalent width measurements, we directly 
compared the synthetic spectra with the observed ones, evaluating the best 
agreement by eye. The microturbulence velocity was fixed to $\xi=1$\,\kms\ 
throughout, which meant that weak lines in some cases gave different 
abundances from stronger lines. However, by combining several lines of each 
element an ionization agreement within $0.2-0.5$\,dex (1\,$\sigma$) was 
reached. This error is comparable to that introduced by not
using the appropriate values of surface gravity in each stellar
case and is acceptable for the present characterisation of the stars
to demonstrate abundance variations.  Comparison
with the relative estimates  for HD\,92499 by \citet{hubrigetal07}
shows reasonable agreement within the errors (differences may
be explained by a possible inhomogeneous chemical surface distribution).

Further limitations of our simple analysis 
include: sensitivity of the derived abundances to errors of
the photometric temperature estimates (we find, e.g., that $\pm250$\,K 
may correspond to 2--33 per cent differences in equivalent width for 
the elements in Table\,\ref{tab:abundance}, worst for Pr); and choice of 
microturbulence $\xi$ for which a change 
of 1\,\kms\ may infer a change in abundances comparable to their
stated errors (especially Nd, see, e.g., \citealt{cowleyetal00}).

Conventionally, microturbulent motions are those with 
characteristic lengths that  are small compared to the mean free path of 
a photon. The value of $\xi$ is normally increased to simulate the effect of
a strong magnetic field (magnetic intensification), but also chemical 
stratification (which may dominate 
the errors of our relative abundances by increasing the differences in 
abundance yields from weak and strong lines) produces effects adverse 
to microturbulence. 
Because of horizontal inhomogeneities and vertical
stratification of the studied sample of magnetic Ap stars, abundance analysis 
is a demanding task (see examples of such studies in the Introduction)
and is outside the aim of the present discovery paper.
We are here interested in relative abundances
to verify peculiar abundances and different abundances between
elements of different ionization states. For this purpose, the estimated
abundance errors of $0.2-0.5$ dex are acceptable.

Table\,\ref{tab:abundance} gives the relative abundances found, and
comments on the individual stars are given in Sect.\,\ref{sec:individual}.
We used 30--70 lines of the 5 elements per star, but the actual selection
of lines varied from star to star.
Note that the models used for HD\,135728 had the simplifying
assumption of a combined continuum of two similar stars (with the 
same luminosity and spectral type). In their study of spectroscopic signatures 
of roAp stars, \citet{ryabchikovaetal04} note the following Pr and Nd anomalies,
expressed as differences in $\log({N/N}_{\rm tot})$ of second and first ionized
states,  
for roAp stars (further including the recent roAp discoveries HD\,116114 and 
HD\,137909) $\Delta[{\rm Pr}]_{\rm III-II}=-0.22$ to 2.19 and 
$\Delta[{\rm Nd}]_{\rm III-II}=0.14$  to 2.55. For other Ap stars, the 
anomalies appear smaller with corresponding ranges 
$\Delta[{\rm Pr}]_{\rm III-II}=-0.51$ to 1.38 and 
$\Delta[{\rm Nd}]_{\rm III-II}=-0.36$ to 1.98. The authors thus suggest that 
roAp stars may have abundance differences for the first two ionized states
of Pr and Nd of at least 1.5 dex and up to 2.5 dex, while non-pulsating stars show 
marginal differences. 
Only 4 stars in Table\,\ref{tab:abundance} had clear indications of ionization 
disequilibria for both
Pr and Nd. These stars are HD\,44226, HD\,92499 (as already found by 
\citealt{hubrigetal07}), HD\,96237 and HD\,143487 and they are therefore promising
roAp star candidates.  Furthermore, HD\,33629, HD\,42075 and HD\,69013 have
significantly different abundances for \ion{Nd}{ii} and \ion{Nd}{iii},
though the same is not clear for Pr. 
\ion{Si}{ii}  6347\,\AA\ and 6371\,\AA\ are strong 
in all spectra, and even prominent in most except for HD\,69013, HD\,96237 and 
HD\,143487. 

\begin{table*}
\caption{\label{tab:abundance}
Relative abundance estimates with number of lines used indicated (n). Estimated 
errors are $0.2-0.5$\,dex. Columns 11--12 give the abundance differences
between the first two ionized states of Pr and Nd; significant ionization
disequilibrium anomalies are indicated with bold font. Last column is the Michigan 
classification \citep{houk78,houk82,houketal75,houketal88}.
The last row is for the solar atmosphere \citep{asplundetal05}.}
\begin{tabular}{@{\,}l@{\,}r@{~}c@{~}r@{~}c@{~}r@{~}c@{~}r@{~}c@{~}r@{~}c@{~}r@{~}
c@{~}r@{~}c@{~}r@{~}c@{~}r@{~}c@{}r@{~}r@{}c@{}
} \hline
     &\multicolumn{20}{c}{{$\log({N/N}_{\rm tot})$}}  \\
\vspace{-0.2cm}\\

HD &              
\ion{Cr}{i}&n&\ion{Cr}{ii}&n&\ion{Fe}{i}&n&\ion{Fe}{ii}&n&\ion{Pr}{ii}&n&\ion{Pr}{
iii}&n&\ion {Nd}{ii}&n&\ion{Nd}{iii}&n&\ion{Eu}{ii}&n&$\Delta[{\rm Pr}]_{\rm III-
II}$&$\Delta[{\rm Nd}]_{\rm III-II}$&Class.\\ \hline
33629  & $ -5.60$      & 3 &$-5.44$& 5   &$-4.29$& 16 &$-4.37$ & 6  & $  -8.57 $ & 
4 & $ -8.80$ & 3 & $ -8.96$ & 8  & $ -7.93$ & 3 & $ -9.68$ & 3 & $-0.23$~~&$  
{\bf 1.03}~~$&{\it *Ap SrCr(Eu)}\\   
42075  & $ -5.83$      & 4 &$-5.68$& 6   &$-4.55$& 17 &$-4.24$ & 7  & $  -8.27 $ & 
3 & $ -8.23$ & 4 & $ -8.30$ & 11 & $ -7.00$ & 2 & $ -8.50$ & 3 & $ 0.04$~~&$  
{\bf 1.30}~~$&{\it *Ap\,EuCrSr}\\ 
44226  & $<$$-5 $      & 1 &$-4.53$& 4   &$-3.79$& 8  &$-4.09$ & 7  & $<$$-9.8 $ & 
2 & $ -8.20$ & 3 & $ -8.37$ & 4  & $ -7.53$ & 3 & $ -8.95$ & 2 & $>${\bf 1.60}~~&$  
{\bf 0.84}~~$&{\it *Ap\,SrEuCr}\\
46665  & $ -5.55$      & 6 &$-5.18$& 8   &$-4.99$& 15 &$-4.43$ & 6  & $  -8.80 $ & 
2 & $ -8.80$ & 3 & $ -8.17$ & 17 & $ -7.83$ & 2 & $ -9.27$ & 3 & $ 0.00$~~&$  
0.34~~$&{\it *Ap\,EuSrCr}\\
47009  & $ -5.55$      & 4 &$-5.73$& 4   &$-4.76$& 7  &$-4.52$ & 6  & $  -8.40 $ & 
3 & $ -9.13$ & 3 & $ -8.23$ & 9  & $ -8.20$ & 3 & $ -8.73$ & 3 & ${\bf -0.73}$~~&$  
0.03~~$&{\it *Ap\,EuCr(Sr)}\\ 
52847  & $ -3.60$      & 4 &$-3.48$& 12  &$-4.24$& 14 &$-3.57$ & 5  & $  -9.13 $ & 
4 & $ -9.25$ & 2 & $ -8.30$ & 6  & $ -8.30$ & 2 & $ -7.57$ & 3 & $-0.12$~~&$  
0.00~~$&{\it *Ap\,CrEu(Sr)}\\ 
55540  & $ -4.63$      & 3 &$-3.92$& 13  &$-5.02$& 6  &$-4.34$ & 7  & $  -8.43 $ & 
3 & $ -8.48$ & 5 & $ -8.23$ & 4  & $ -8.30$ & 2 & $ -8.90$ & 3 & $-0.05$~~&$ -
0.07~~$&{\it *Ap\,EuCr}\\ 
69013  & $ -6.10$      & 3 &$-5.50$& 5   &$-4.71$& 15 &$-4.70$ & 5  & $  -7.45 $ & 
3 & $ -8.33$ & 3 & $ -8.34$ & 21 & $ -7.10$ & 3 & $ -9.15$ & 3 & ${\bf -0.88}$~~&$  
{\bf 1.24}~~$&{\it *Ap\,EuSr}\\
72316  & $ -4.83$      & 4 &$-4.70$& 13  &$-4.59$& 16 &$-4.09$ & 9  & $  -9.90 $ & 
2 & $ -9.45$ & 4 & $ -8.55$ & 7  & $ -8.65$ & 2 & $ -9.23$ & 3 & $ 0.45$~~&$ -
0.10~~$&{\it *Ap\,EuCr(Sr)}\\ 
75049  & $ -4.70$      & 1 &$-4.56$& 7   &$-4.30$& 5  &$-4.04$ & 11 & $  -8.50 $ & 
1 & $ -8.25$ & 4 & $ -7.65$ & 2  & $ -7.45$ & 2 & $ -7.80$ & 3 & $ 0.25$~~&$  
0.20~~$&{\it *Ap\,EuCr}\\ 
88701  & $ -4.81$      & 8 &$-4.34$& 20  &$-4.96$& 16 &$-4.01$ & 8  & $  -8.50 $ & 
3 & $ -8.95$ & 6 & $ -8.62$ & 9  & $ -8.75$ & 2 & $ -8.90$ & 3 & $-0.45$~~&$ -
0.13~~$&{\it *Ap\,CrSi}\\
92499  & $ -5.90$      & 4 &$-4.93$& 3   &$-4.59$& 12 &$-4.35$ & 6  & $  -8.70 $ & 
3 & $ -8.20$ & 3 & $ -8.08$ & 18 & $ -6.83$ & 4 & $ -9.30$ & 3 & ${\bf  0.50}$~~&$  
{\bf 1.25}~~$&{\it *Ap\,SrEuCr}\\ 
$96237_{\phi1}$&$-5.84$& 5 &$-5.10$& 6   &$-4.71$& 9  &$-4.71$ & 5  & $  -8.28 $ & 
5 & $ -7.64$ & 5 & $ -6.80$ & 17 & $ -5.60$ & 3 & $ -9.85$ & 3 & ${\bf  0.64}$~~&$  
{\bf 1.20}~~$&{\it *Ap\,SrEuCr}\\ 
$96237_{\phi2}$&$-4.98$& 4 &$-4.72$& 5   &$-4.24$& 8  &$-4.15$ & 3  & $  -9.55 $ & 
2 & $ -8.23$ & 4 & $ -8.72$ & 5  & $ -6.43$ & 3 & $ -8.73$ & 3 & ${\bf  1.32}$~~&$  
{\bf 2.29}~~$&\\ 
110274 & $ -4.89$      & 7 &$-4.58$& 19  &$-4.57$& 17 &$-4.25$ & 10 & $  -9.23 $ & 
4 & $ -9.02$ & 5 & $ -8.77$ & 13 & $ -8.20$ & 2 & $ -8.63$ & 3 & $ 0.21$~~&$  
{\bf 0.57}~~$&{\it *Ap\,EuCr}\\ 
117290 & $ -5.08$      & 6 &$-4.78$& 11  &$-4.33$& 17 &$-4.28$ & 13 & $  -9.18 $ & 
4 & $ -9.30$ & 3 & $ -8.89$ & 10 & $ -8.80$ & 5 & $ -8.67$ & 3 & $-0.12$~~&$  
0.09~~$&{\it *Ap\,EuCrSr}\\ 
121661 & $ -5.10$      & 3 &$-4.25$& 5   &$-4.52$& 10 &$-3.98$ & 4  & $  -8.97 $ & 
3 & $ -8.78$ & 3 & $ -8.14$ & 7  & $ -7.88$ & 2 & $ -8.97$ & 3 & $ 0.19$~~&$  
0.26~~$&{\it *Ap\,EuCr(Sr)}\\ 
135728A&               &   &$-5.67$& 4   &$-3.36$& 5  &$-3.35$ & 6  &$<$$-10.50$ & 
2 & $-10.20$ & 2 & $ -8.70$ & 4  & $ -8.70$ & 2 & $ -9.40$ & 2 &$>$0.23~~&$ -
0.24~~$&{\it *Ap\,SrEuCr}\\ 
135728B& $ -4.48$      & 3 &$-4.47$& 9   &$-4.30$& 5  &$-4.30$ & 3  & $  -8.83 $ & 
3 & $ -9.13$ & 4 & $ -8.31$ & 8  & $ -8.10$ & 2 & $ -8.80$ & 2 & $ 0.07$~~&$ -
0.03~~$&\\ 
143487 & $ -5.50$      & 3 &$-5.33$& 8   &$-5.09$& 9  &$-4.70$ & 9  & $  -9.20 $ & 
3 & $ -7.70$ & 2 & $ -8.08$ & 11 & $ -6.72$ & 5 & $ -8.60$ & 2 & ${\bf  1.50}$~~&$  
{\bf 1.36}~~$&{\it *APEC}\\  
Sun    & $ -6.40$      &   &$-6.40$&     &$-4.59$&    &$-4.59$ &    & $ -11.33 $ &   
& $-11.33$ &   & $-10.59$ &    & $-10.59$ &   & $-11.52$ &   &        &    &   \\
\hline \hline
\end{tabular}
\end{table*}

\subsubsection{Magnetic field measurements}
\label{sec:mag}

A strong field, typically $\sim 2-3$\,kG, combined with a slow projected rotation
rate (smaller than \vsini\ $\sim\,10$\,\kms) can produce magnetically resolved 
lines by the Zeeman effect \citep{mathysetal97}.
In the simplest cases of spectral lines corresponding to doublet or
triplet Zeeman patterns, simple formulae can be applied to determine
in a virtually approximation-free manner the mean magnetic field
modulus $\langle B\rangle$ from measurement of the wavelength
separation of the resolved Zeeman components \citep{mathys89}. 
$\left<B\right>$ is the average of the modulus of the magnetic vector, over 
the visible stellar hemisphere, weighted by the local line intensity. For a
triplet pattern, its value (in G) is obtained from the wavelength
separation $\Delta\lambda$, between the central $\pi$ component and either of the
$\sigma$ components, by using the formula:
\begin{equation}
  \centering
  \left<B\right>=\Delta\lambda/(4.67\cdot10^{-13}\;\lambda_{c}^{2}\;g_{\rm eff}),
  \label{eq:zeeman1}
\end{equation}
where $\lambda_{\rm c}$ is the central wavelength of the line 
and $g_{\rm eff}$ is the effective Land\'e factor of the transition.
Both $\Delta\lambda$ and $\lambda_{\rm c}$ are expressed in \AA. 
For a doublet pattern, the relation between $\langle B\rangle$ and 
the separation $\Delta\lambda$ of the split components (each of which is the
superposition of a $\pi$ and a $\sigma$ component) is:
\begin{equation}
  \centering
  \left<B\right>=\Delta\lambda/(9.34\cdot10^{-13}\;\lambda_{c}^{2}\;g_{\rm eff}).
  \label{eq:zeeman}
\end{equation}

\begin{figure*}
 \vspace{-3pt}
 \hspace{-13pt}
\includegraphics[height=0.71\textheight, width=0.79\textwidth, angle=0]{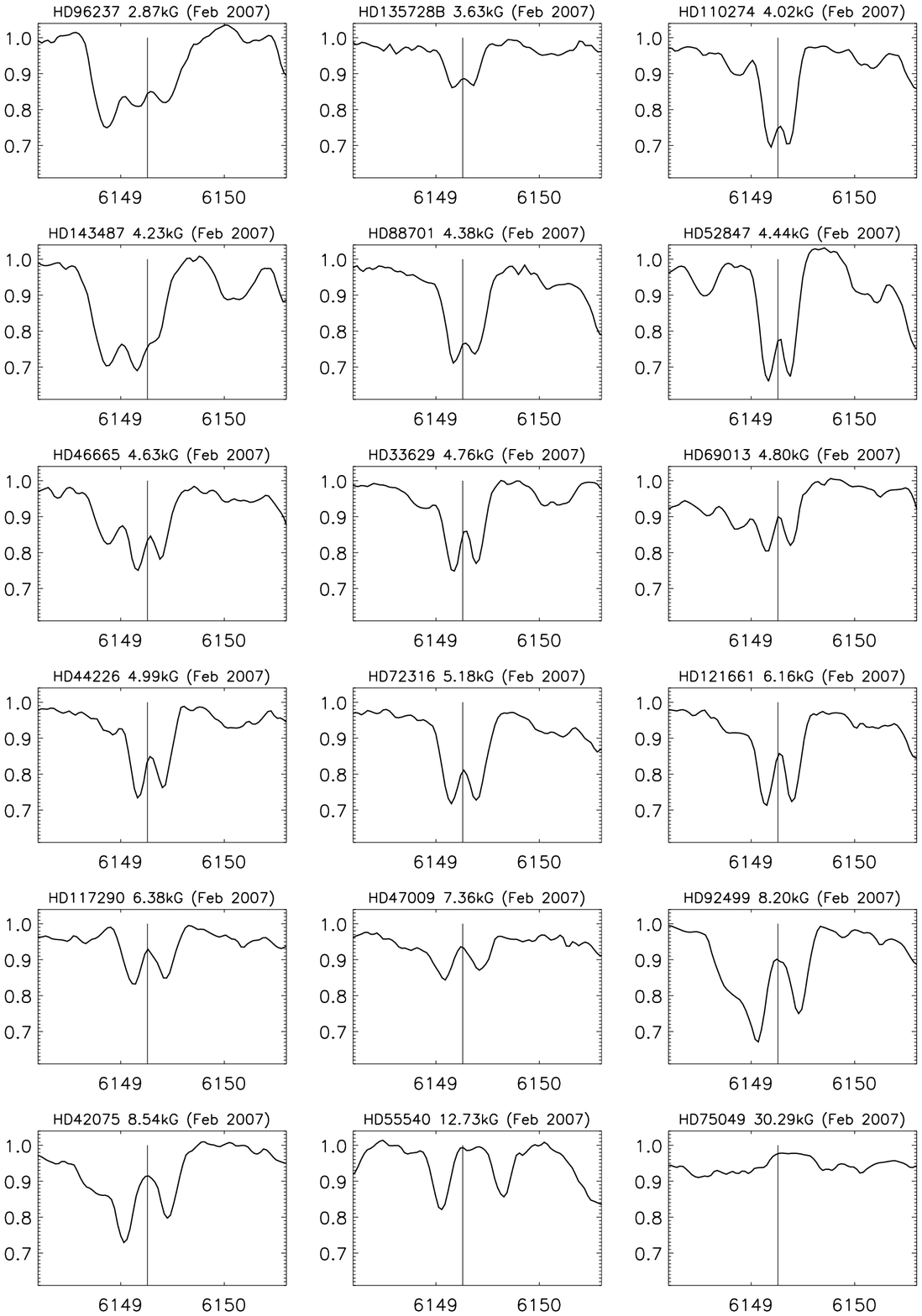}
 \caption{\label{fig:fe6149a}Comparison of the magnetically 
 resolved \ion{Fe}{ii}\,6149\,\AA\  lines in the 
2007 February (FEROS) spectra. 
 The abscissa is wavelength in \AA, the ordinate is 
 normalized intensity.
 Radial velocity shifts were added to the laboratory wavelengths
 estimated from \halpha. Note that for HD\,96237, the region with
 $\lambda$6149\,\AA\  is dominated by unidentified rare earth elements as iron
 is particularly weak at this rotation phase (Table\,\ref{tab:abundance}). The 
spectra appear from top (left) to bottom (right) with increasing magnetic 
field modulus $\left<B\right>$ from Table\,\ref{tab:targets}.
}
\end{figure*}

\begin{figure*}
 \vspace{-3pt}
 \hspace{-13pt}
\includegraphics[width=0.73\textwidth]{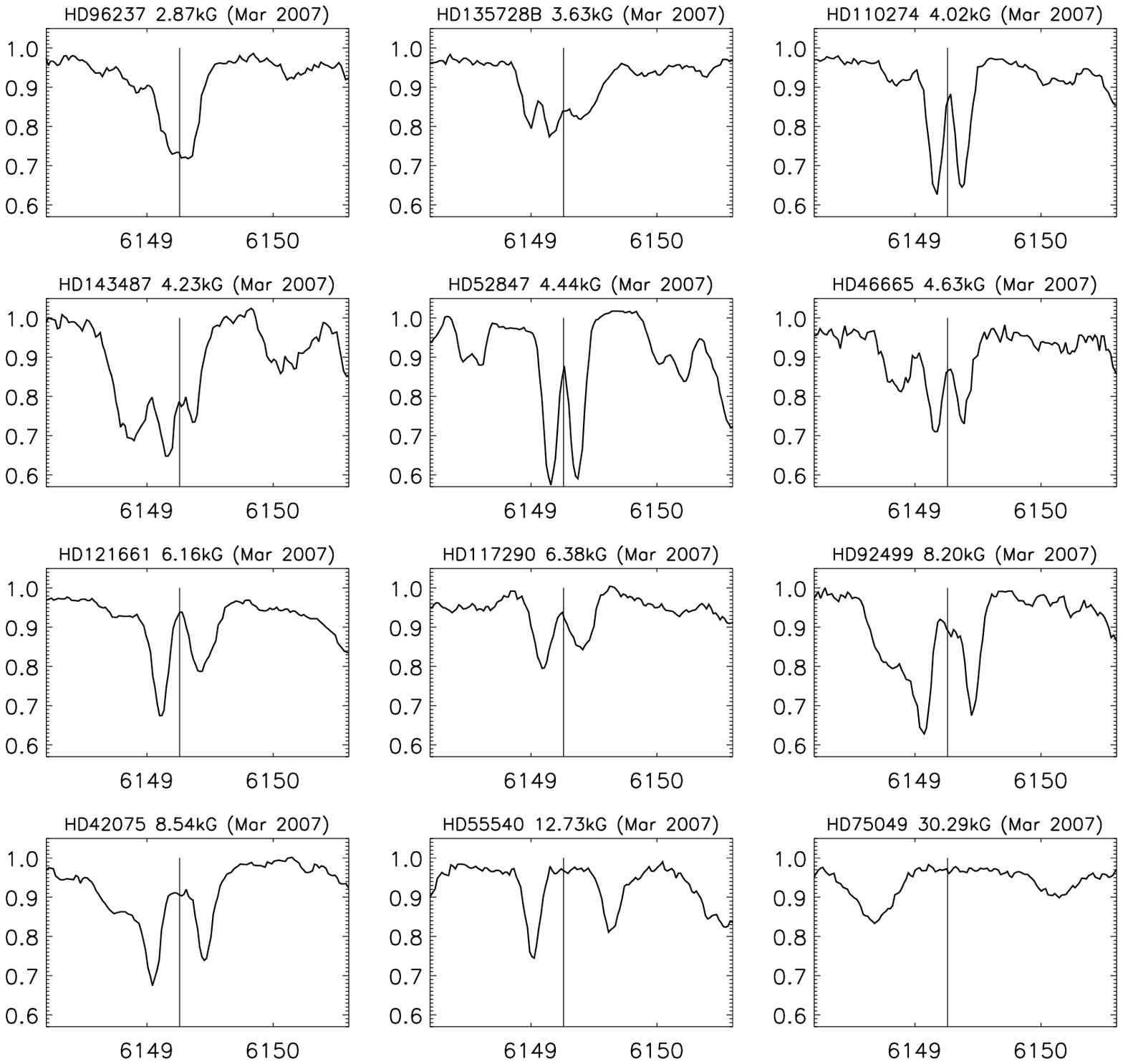}
 \caption{\label{fig:fe6149b}Same as Fig.\,\ref{fig:fe6149a}, but 
for UVES spectra obtained about 1 month later for some of the stars.}
\end{figure*}

As shown in Fig.\,\ref{fig:fe6149a}, the Zeeman splitting
of the $\lambda$6149\,\AA\ line is visible for nearly all 18 stars in the FEROS
spectra (HD\,96237 excepted, see Sect.\,\ref{sec:96237}).
The UVES follow-up observations  (Fig.\,\ref{fig:fe6149b}) 
showed considerable changes in this line for 4 stars: 
{\it HD\,143487} which has only partially split lines in the FEROS spectrum; 
{\it HD\,135728AB}, an SB2 binary; 
{\it HD\,96237} which is highly variable and peculiar, and which showed magnetically
resolved lines in the 2007 March spectrum only, and then only for a few Cr and Fe
lines. The $\lambda$6149\,\AA\ line in this star is never seen as double, 
but appears broadened
in the 2007 March observations (Fig.\,\ref{fig:fe6149b}) to an extent
that is in good agreement with a magnetic field modulus of 2.6\,kG; and
finally {\it HD\,75049} which is an extremely magnetic star that shows, 
probably due to rotation, a changing magnetic field strength and 
spectral fine-structure, such as appearance and disappearance of the Paschen-Back 
effect in certain wavelength regions. 
The mean magnetic field modulus $\langle B\rangle$ measurements based on
$\lambda$6149\,\AA\ are given in Table\,\ref{tab:targets}. The corresponding
errors are based on measurements repeated in different spectra when available.
These errors may be affected by spectrum-to-spectrum  variations in the 
magnetic field strength itself, related to the stellar rotation. Errors are 
also sensitive to conditions such as \vsini, blends from close lines, and 
the magnetic field strength (i.e. how well separated the Zeeman components
are). For 6 stars with only
a single spectrum available, the errors were estimated based on \vsini, and
on errors found for stars with similar $\lambda$6149\,\AA\ line profile 
shapes, using as a lower limit the scatter
from repeating the measurements using different settings for treating blends and 
continuum placement.  Further details are given in Sect.\,\ref{sec:individual}.


Measurements of line broadening and of separation of 
Zeeman split components in magnetically 
resolved lines were made by using  {\small IRAF}'s {\small onedspec.splot}
task. For blended cases,   $2-4$ Gaussian profiles were fitted 
simultaneously and Eqs.\,\ref{eq:zeeman1} and \ref{eq:zeeman} were
applied to determine the mean magnetic modulus $\left<B\right>$. 
The magnetic measurements were made mostly for the same set of lines, but
due to differences in line broadening, magnetic splitting, differences in
abundances resulting in different degrees of line blending, the
choice of lines used 
was always made on a star-by-star basis.  The most important lines used,
with large Land\'e factors ($g_{\rm eff} = 1.2-2.9$), are:
\ion{Cr}{i}\,5247.56\,\AA,
\ion{Cr}{ii}\,5116.04 and 6112.26\,\AA,
\ion{Eu}{ii}\,6437.64\,\AA,
\ion{Fe}{i}\,5068.76, 6230.72, 6232.64, 5324.17 and 6336.82\,\AA,
and \ion{Fe}{ii}\, 6149.25, 6369.46, 6383.72, 6446.41 and 6600.02\,\AA.

Spectra of the stars re-observed in 2007 March were used for 
checking radial velocity (RV) stability by measuring the 
two non-magnetically sensitive iron lines \ion{Fe}{i}\,5434.52\,\AA,
\ion{Fe}{ii}\,6586.69\,\AA\  and occasionally also \ion{Nd}{iii}\,6145.63\,\AA. 
The wavelength scales in all cases have been
added offsets determined from lines selected in the telluric line 
list by \citet{griffinetal73}. This ensures accurate 
radial velocities to the level of $300\pm200$\,\ms, sufficient for
first checks on duplicity for the considered interval of $30-35$\,d.

\subsection{Individual notes}
\label{sec:individual}

In this section individual magnetic stars 
are described in sequence of appearance in the Henry Draper Catalogue. 
Spectral classifications from the Michigan 
Spectral Catalogue 
are given in Table\,\ref{tab:abundance} and repeated here when
individual notes from the Catalogue were available.

To supplement the $\lambda$6149\,\AA\ measurements in 
Table\,\ref{tab:targets}, Zeeman measurements for other lines
are also given. However, while $\lambda$6149\,\AA\ is a
heavily used diagnostic line for which the blending is rather well known,
interpretation of the field strengths measured for other lines, 
especially of different chemical elements, is not straightforward.
This in part because the Land\'e factors of different lines not
always have the same reliability, but more importantly because 
blending of other (non-diagnostic) lines may be much less secure. 
Furthermore, lines of elements such as Cr or REEs may have
differences in their distribution over the stellar surface, meaning that the
field measurement is weighed differently. 
Differences between field values from lines of different elements thus result 
in a scatter that is larger than for $\lambda$6149\,\AA\ alone based on more 
spectra. Most likely it reflects actual physical effects in the star, rather 
than provides a reliable estimate of the measurement errors.

\subsubsection{HD\,33629}
From the magnetic components of $\lambda$6149\,\AA\ we measure a magnetic field of 
$4.76\pm0.17$\,kG, which combined with the measurements of 
\ion{Fe}{i}\,6232.64\,\AA\  and \ion{Fe}{ii}\,6369.46\,\AA\ 
 gives  $\left<B\right>=4.21\pm0.48$\,kG. 
The photometric temperature is $\teff=7570$\,K.
The abundance estimates in Table\,\ref{tab:abundance} show 
overabundances of Nd, Pr, Eu and Cr compared to the Sun, a solar Fe abundance, and 
ionization disequilibrium for Nd only.
The line \ion{Ba}{ii}\,6141.71\,\AA, common in many roAp stars, is very strong.
The ASAS photometry excludes  rotational photometric variability 
greater than 5.1\,mmag amplitude. 

\subsubsection{HD\,42075}
Observed on three different nights (see Table\,\ref{tab:obslog}), this star 
had a constant radial velocity  $4.2\pm0.4$\,\kms. From the resolved 
$\lambda$6149\,\AA\ 
line, all spectra give $\left<B\right>=8.54\pm0.02$\,kG which with Zeeman 
splitting
measured for 6 additional iron lines shows a stable magnetic field
of $8.58\pm0.14$\,kG for all nights combined. 
ASAS photometry indicates no variability above
4.0\,mmag. The star's sharp-lined spectrum (\vsini\,$\sim 1$\,\kms) adds further 
support
to a slow rotation period, longer than a  month.
The abundance estimates in Table\,\ref{tab:abundance} show
overabundances of Cr, Pr, Nd and Eu, and near-solar Fe. 
 \ion{Ba}{ii}\,6141.71\,\AA\  is very strong (see also Figs.\,\ref{fig:abunall} 
and \ref{fig:abunhal}).

\subsubsection{HD\,44226}

The $\lambda$6149\,\AA\ line indicates $\left<B\right>=4.99\pm0.01$\,kG  which
with \ion{Fe}{i}\,6232.64\,\AA\  and \ion{Fe}{ii}\,6369.46\,\AA\  combined gives
 $\left<B\right>=5.04\pm0.52$\,kG. 
ASAS photometry excludes periodic variability above 5.3\,mmag.
The spectrum is  similar to that of HD\,33629 and also shows
a very strong  \ion{Ba}{ii}\,6141.71\,\AA\ line. All abundances in
Table\,\ref{tab:abundance} are greater than solar. As ionization
disequilibria for both Nd and Pr are
similar to those of known roAp stars, 
$\Delta[{\rm Pr}]_{\rm III-II}\ge1.60$ and $\Delta[{\rm Nd}]_{\rm III-II}=0.84$,
this star is a good roAp candidate.

\subsubsection{HD\,46665} 

This star shows no long-period photometric variability above 5.1\,mmag in $V$
(from the ASAS survey).
Our two spectra obtained 35\,d apart indicate an unchanged radial velocity, 
$22.9\pm0.2$\,\kms.
The magnetic measurements for $\lambda$6149\,\AA\  give similar values,
4.72 and 4.53\,kG, or $\left<B\right>=4.63\pm0.13$\,kG combined. This is 
supported by including an 
additional $2-3$ Cr and Fe lines which give 
$4.64\pm0.23$ and $4.68\pm0.13$\,kG for the FEROS and UVES spectra 
respectively.  The
spectrum shows an absence of barium (especially $\lambda$6141\,\AA)
and lines of silicon are weak. Overabundances of Cr, Nd, Pr and Eu 
are found, while the Fe abundance is solar.  The
 Michigan Catalogue classification and note is  
{\it *Ap\,EuSrCr, or possibly Si rather than Eu; this is the strongest of the 
metal lines,} but note that we find weak Si in our recent spectra.  

\subsubsection{HD\,47009}    
The  $\lambda$6149\,\AA\ line indicates a  
field strength of $7.36\pm0.13$\,kG, confirmed by adding measurements for
two more Cr and Fe lines which gives $7.49\pm0.19$\,kG.
Photometrically, the star is stable to 4.1\,mmag in $V$.  The temperature is at the
high end of our sample, $\teff=8280$\,K, as supported by the spectra.
Long stretches of continuum are seen. Ba is absent, while Si and Ca are
clearly present. Cr, Nd, Pr and Eu are greater than solar, while
Fe is solar.

\subsubsection{HD\,52847} 

\citet{hensbergeetal84} found this star to be photometrically stable in
Str\"omgren $u$ to less than 15\,mmag, which the ASAS $V$ photometry 
confirms to 4.2\,mmag. 
We obtained two spectra of the star, 35\,d apart, both with the
same radial velocity within the errors, 11.3\,\kms.
The \ion{Fe}{ii}\,6149\,\AA\  line gives a constant magnetic field of 
$4.44\pm0.01$\,kG for both spectra, while adding $2-5$ more Cr, Fe and Eu lines
to the analysis confirms this result as $\left<B\right>=4.76\pm0.61$ 
(HJD\,2454138) and $\left<B\right>=4.48\pm0.15$\,kG  (HJD\,2454173).
Note the considerable error compared to that of the 
{\em diagnostic} \ion{Fe}{ii}\,6149\,\AA\  line,
because of the effects mentioned in the beginning of Sect.\,\ref{sec:individual}.
Of the studied stars, we find this star has the highest abundance of Cr,
while Fe is near-solar and Nd, Pr and Eu are greater than solar.
Ca is weak (\ion{Ca}{ii}\,K is sharp, corresponding to spectral type early A). 
Ba $\lambda$6141\,\AA\  is strong and Eu has the highest abundance of the studied 
stars. 

\subsubsection{HD\,55540} 
Based on 15 single nightly measurements obtained over some years, 
\citet{hensbergeetal84} claimed this star
to be a long period variable with a 
44\,mmag variability in the Str\"omgren $u$ band, at the 2.5-$\sigma$ level.
However, the ASAS $V$ band shows stability to 5.2\,mmag over 5.6 years, at
which amplitude a candidate period of 42.1\,d is found. The significance
is marginal, 4.2\,$\sigma$, but without {\it Hipparcos} photometry it is not
a reliable detection.  As a strong magnetic field was obvious in this star's 
spectrum (12.5\,kG),  we observed it on 4 different nights in 2007 February and 
March. The three 2007 February spectra show similar radial velocity, 
${\rm RV}=49.9-50.5$\,\kms, while the 2007 March spectrum exhibits a 
significantly lower velocity, ${\rm RV}=36.6$\,\kms. 
With typical radial velocity  errors of $0.5-0.8$\,\kms, this is an unconfirmed 
detection of a radial velocity variable and probable SB1 binary. 
No indication of a secondary spectrum was noted.
From Zeeman splitting measurements based on $5-7$ Cr, Nd, Eu and Fe lines,
we find that the magnetic field is slightly, but significantly,
increasing: 
$\left<B\right>=12.38\pm0.13$ (HJD\,2454138),
$\left<B\right>=12.29\pm0.26$ (HJD\,2454139),
$\left<B\right>=12.40\pm0.21$ (HJD\,2454141) and
$\left<B\right>=13.04\pm0.21$\,kG (HJD\,2454173).
The corresponding individual \ion{Fe}{ii}\,6149\,\AA\  measurements 
give respectively $\left<B\right>=12.50$, 12.50, 12.63 and 
13.07\,kG ($\left<B\right>=12.73\pm0.30$\,kG, 
combined).
Cr, Nd, Pr and Eu are greater than solar and Fe is near-solar, and
Ba $\lambda$6141\,\AA\ is present.
The temperature (8280\,K) is high for known roAp stars.

\subsubsection{HD\,69013}
The $\lambda$6149\,\AA\ line  indicates a magnetic 
field strength of $4.80\pm0.06$\,kG. When combining this line with two more iron 
lines and \ion{Eu}{ii}\,6437.64\,\AA\  (only partially split), a field of 
$\left<B\right>=4.33\pm0.39$\,kG is determined.
There is no variability seen with the ASAS photometry to 4.1\,mmag for this 
relatively faint star ($V=9\fm45$). 
Cr and Fe abundances are near-solar and solar, respectively, while abundances of
Nd and Pr are some of the highest in this study. Eu is also greater than solar.
The Ba, Ca and Si lines are all prominent.
Nd and Pr ionization disequilibrium anomalies of  $\sim 1$\,dex are found
in opposite directions, which is not seen in known roAp stars.

\subsubsection{HD\,72316}    
Based on $\lambda$6149\,\AA, a field of $\left<B\right>=5.18\pm0.33$\,kG 
is found. This is confirmed 
with the \ion{Cr}{ii} lines $\lambda\lambda$\,5116.05\,\AA\  and 5318.38\,\AA\ 
which indicate $\left<B\right>=5-6$\,kG.
ASAS photometry shows no variability
above 
4.5\,mmag.
\ion{Si}{i} and
Ca are weak and barely visible. A sharp \ion{Ca}{ii}\,K line supports the
photometric temperature of $\teff=8050$\,K (early to mid-A type). 
Ba $\lambda$6141\,\AA\  is present ($\sim 4$ per cent absorption below continuum).
All Cr, Fe, Nd, Pr and Eu abundances are greater than solar, but Nd, Pr and Eu 
have some of the lowest abundances in the studied sample of 
stars.  In the astrometric H-R 
diagram, the star appears to be near the end of its main sequence lifetime.

\subsubsection{HD\,75049}
\begin{figure}
\includegraphics[height=0.48\textwidth, angle=90]{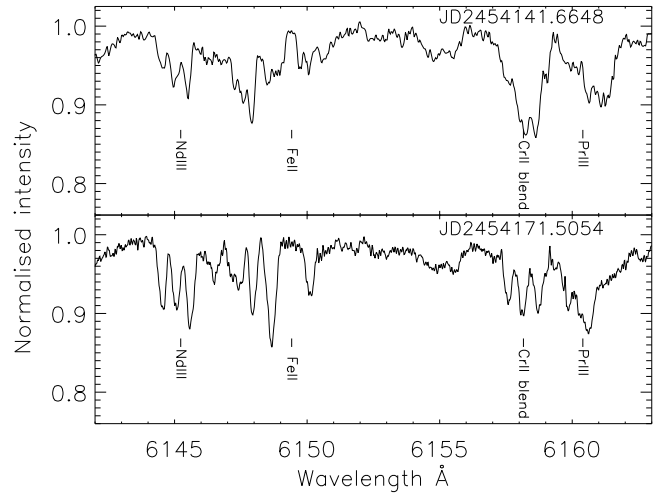}
 \caption{\label{fig:HD75049a}
A portion of the spectrum of HD\,75049 recorded with FEROS on 
2007 February 10 (top), 
and with UVES on 2007 March 11 (bottom). 
Very significant variations of the spectral 
line intensities and shapes between the two epochs are clearly seen.}
\end{figure}

This star has the second strongest magnetic field of all known Ap 
stars, and if this field is variable, it may even surpass that of
HD\,215441 (34.4\,kG, \citealt{babcock60}) at some phases. Our
spectra, separated by 30\,d, show considerable differences in the
field strength and component fine-structure (Fig.\,\ref{fig:HD75049a})
probably related with viewing different aspects of the star's
magnetic field and surface element distribution.
The strength of the field is such that the splitting
pattern of the $\lambda$6149\,\AA\ line is strongly distorted by
partial Paschen-Back effect, 
hampering its use as a diagnostic line.
The splitting observed in the 2007 February spectrum cannot be simply
interpreted, while the appearance of the line in the 2007 March spectrum is
more similar to the doublet pattern observed in stars with
weaker fields.
If the separation of the two components is translated to
a magnetic field in the usual manner, a value of 
$\left<B\right>=30.29\pm0.08$\,kG is obtained 
for this epoch.  However, application of Eq.~(\ref{eq:zeeman}) in this case is not
quite justified.  Combining the 
\ion{Fe}{ii}\,6149\,\AA\ measurement with those of an additional 4 Nd, Eu and
Fe lines we obtain $\left<B\right>=30.33\pm0.48$\,kG.

The 2007 February spectrum's lower resolution, together with the partial Paschen-Back 
effect, limits the number
of resolved lines available for magnetic field measurements. 
With a few lines, \ion{Nd}{iii}\,6145.07\,\AA,
\ion{Nd}{iii}\,6550.32\,\AA\ and \ion{Fe}{ii}\,6456.38\,\AA, a field of
 $\left<B\right>=25.77\pm1.81$\,kG is obtained.
Radial velocity measurements ($10.2\pm0.6$\,\kms, combined) show no differences 
for the two spectra above 1.0\,\kms, which is within the measurement errors.

\begin{figure}
\includegraphics[height=0.46\textheight,width=0.47\textwidth, angle=0]{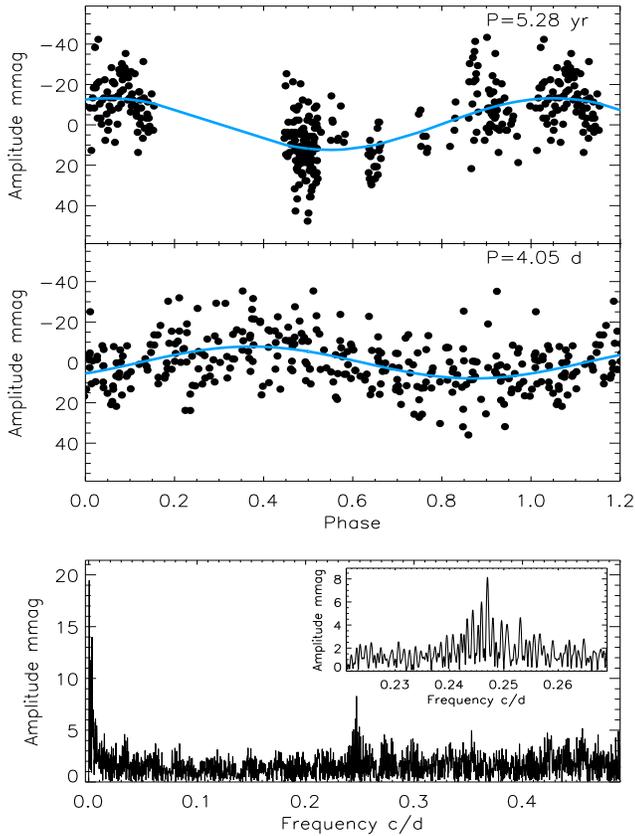}
 \vspace{0.1cm}
 \caption{\label{fig:HD75049asas}
 Upper panel: ASAS light curve of HD\,75049 folded with the dominant 
 5.28\,yr-period. The data were observed in the shown sequence. 
  Middle panel: the same light curve with the 
  5.28\,yr period prewhitened and folded with the 4.05\,d period. 
  Best-fit models are superposed the light curves.
 Lower panel: The corresponding amplitude spectra showing the 5.28\,yr-period
 with the second period at 4.05\,d; the insert is a zoom-in on the latter period,
 after prewhitening for the former.
 Note the smaller amplitude scale.}
\end{figure}

The ASAS photometry reveals (Fig.\,\ref{fig:HD75049asas}) the star to be an 
$\alpha^2$\,CVn rotational variable with two significant 
periodicities:
5.28\,yr (12.7\,mmag amplitude, 7.4\,$\sigma$) and a much
shorter one at 4.05\,d (7.9\,mmag, 5.5\,$\sigma$).
As the spectra show considerable variability in just 30\,d, and
\vsini\,=\,8\,\kms, we suspect that the shorter period may be identified
as the rotation period. The epochs in Table~\ref{tab:obslog} show that our two 
spectra were taken 7.4 rotation cycles apart for the 4.05-d rotation period. This 
is consistent with the observed spectral variability.

From the shape of the \halpha\ wings and from the photometric 
temperature estimate, the temperature of this Ap star is $T_{\rm  eff} = 9700$\,K,
in excellent agreement with the weak and sharp \ion{Ca}{ii}\,K line that indicate
A0 or even late-B type. Ba, \ion{Si}{i} and Ca are weak or 
not visible. (Ba $\lambda$6141\,\AA\  
is present but broadened or scrambled, weak and 
asymmetric). Cr, Nd, Pr and Eu have greater than solar abundances, while Fe is 
solar or near-solar.  No Nd or Pr ionization anomalies are found.
We are currently performing follow-up with spectroscopy and spectropolarimetry to
establish the rotation rate, mean magnetic field modulus and mean longitudinal
field to characterise HD\,75049's magnetic and spectral variations and derive
a first model of its magnetic field.
The spectral classification from the Michigan Catalogue with a note is  {\it 
*Ap\,EuCr. Or very possibly Si rather than Eu.}

\subsubsection{HD\,88701}  

\begin{figure}
\includegraphics[width=0.47\textwidth, angle=0]{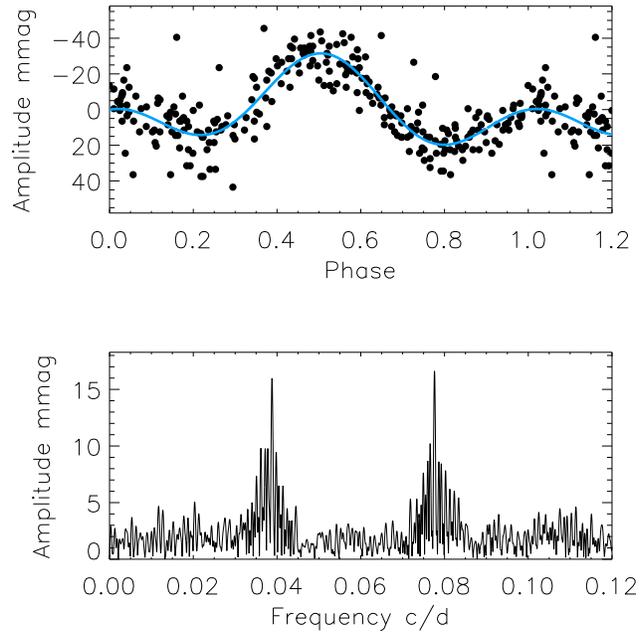}
 \caption{\label{fig:HD88701}
 Top: ASAS light curve of HD\,88701 folded with the rotation period 25.765\,d
 and the best-fit model superimposed. The double-wave is characteristic
 for $\alpha^2$\,CVn variables.
 }
\end{figure}

The $\lambda$6149\,\AA\ 
line indicates a field of $\left<B\right>=4.38\pm0.32$\,kG. With 3 additional Cr 
and Fe lines, we get $\left<B\right>=4.21\pm0.41$\,kG.  Ba, Ca and \ion{Si}{i} 
are weak or absent, while Cr, Pr, Nd and Eu are greater than solar.  Although 
Fe appears to differ in abundances of its two ions, the mean abundance is 
solar.  A
sharp \ion{Ca}{ii}\,K line indicates early-A type, while our photometric
temperature estimate is $\teff=8080$\,K. 
The ASAS $V$ light curve shows
(Fig.\,\ref{fig:HD88701}) a double-wave when phased with the period 25.765\,d. 
This is a clear $\alpha^2$\,CVn signature of a spotted
surface and as \vsini\,=\,7\,\kms\ is 
one of the highest of our sample, 
we identify the period as the rotation period.
If we take the rotation period and measured \vsini\ at face value, the 
implication is that HD\,88701 has a radius of $R = 3.5$\,R$_{\odot}$, an
unexpectedly high value. However, with an error estimate of $\pm 1$\,km\,s$^{-1}$, 
a 3\,$\sigma$ lower limit on \vsini\ of 4\,km\,s$^{-1}$ then leads to a much more 
typical radius for magnetic Ap stars of 2\,R$_{\odot}$. These numbers suggest 
that the rotational inclination is close to $90^{\circ}$.  
The spectral classification from the Michigan Catalogue with a note is  {\it 
*Ap\,CrSi. Fairly weak but definite case.}

\subsubsection{HD\,92499}  

\begin{figure}
 \vspace{4pt} 
 \hspace{1pt}
\includegraphics[height=0.44\textwidth, angle=0]{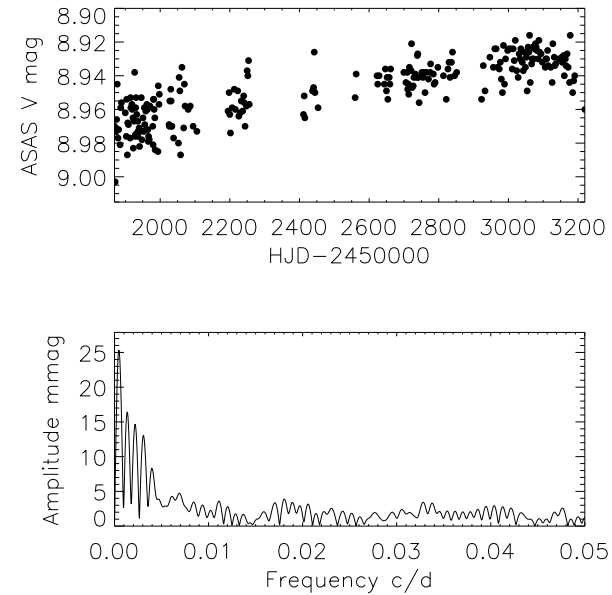}
 \caption{\label{fig:HD92499}
 ASAS light curve of HD\,92499 covering a period of 3.7\,yr.
 Below is the corresponding amplitude spectrum.
 }
\end{figure}

This Ap star was recently discovered to have magnetically resolved 
lines by \citet{hubrigetal07}, who on 2006 May 13 observed the star and 
measured  $\left<B\right>=8.5\pm0.2$\,kG and \vsini\,$=3.0\pm0.5$\,\kms. We 
re-observed the star on 4 different nights. Using 7 Fe lines for the FEROS spectra 
and 20 for the UVES spectrum, we measured:
$\left<B\right>=8.12\pm0.17$ (HJD2454138),
                $8.06\pm0.27$ (HJD2454139),
                $8.17\pm0.12$  (HJD2454141)   and 
                $8.09\pm0.31$\,kG (HJD2454172),
i.e. a constant field strength, slightly smaller 
than Hubrig et al.'s measurement. Of these Fe line measurements, the 
$\lambda$6149\,\AA\ line
alone shows no trends and provides, combined for all 4 spectra, 
$\left<B\right>=8.20\pm0.13$\,kG. Radial velocities of all spectra show 
no variability with  ${\rm RV}=18.39\pm0.25$\,\kms. The ASAS light curve 
(Fig.\,\ref{fig:HD92499}) shows a 40\,mmag increase in 3.7\,yr, while the 
{\it Hipparcos} data (separated from the ASAS data by $\sim 7.5$\,yr)
show no change above 9.8\,mmag in 3.2\,yr, but this 
could occur during light curve maximum or minimum. If the ASAS variability 
is real and related to the rotation, a rotation period longer than 5\,yr 
would be expected.

The abundances of Cr, Pr, Nd and Eu are greater than solar; Fe is near-solar.
Ba $\lambda$6141\,\AA\ is present but very shallow. The \ion{Ca}{ii}\,K line 
suggests 
type late-A to early-F which makes the photometric temperature of $\teff=7810$\,K
seem slightly high.
As found by \citet{hubrigetal07}, we see Pr and Nd ionization anomalies:
$\Delta[{\rm Pr}]_{\rm III-II}$=$0.50$ and $\Delta[{\rm Nd}]_{\rm III-II}$=$1.25$.
This star is a promising roAp candidate.  

\subsubsection{HD\,96237}  
\label{sec:96237}

\begin{figure}
 \vspace{-3pt} 
 \hspace{-28pt}
\includegraphics[height=0.55\textwidth, angle=90]{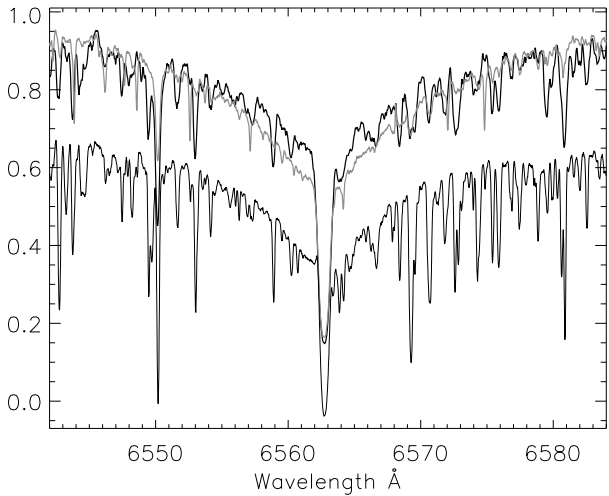}
 \caption{\label{fig:HD96237a}
 Spectral \halpha\ region for HD\,96237  and
 the most peculiar known Ap star, HD\,101065 (spectrum with
 thin line, shifted 0.25 downwards in intensity). 
 The two upper HD\,96237 spectra are from February 
 (black thick line) and March (grey line). Ordinate is normalized intensity.
}
\end{figure}

\begin{figure*}
 \vspace{1pt} 
 \hspace{-18pt}
\includegraphics[height=0.85\textwidth, angle=90]{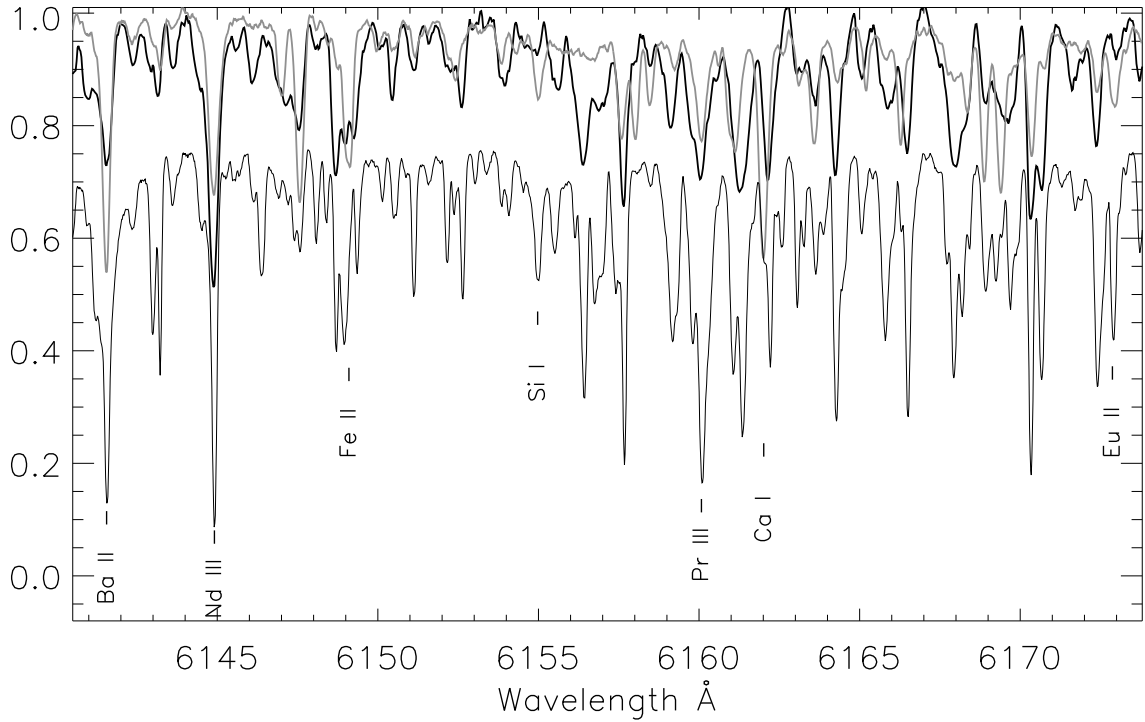}
 \caption{\label{fig:HD96237b} Same as Fig.\,\ref{fig:HD96237a}, 
 but for a region rich in elements such as Nd, Zr, Ce, Ba, La.
 The two HD\,96237 spectra are from February (black thick line) and
 March (grey line). Note the extreme change from a highly
 peculiar spectrum to a typical Ap one, as well as disappearance 
 and intensity changes in several lines. 
}
\end{figure*}

This star is enigmatic and intriguing. It is best described as a highly
peculiar, magnetic Ap star with a spectrum that changes dramatically in the 
1-month time span of our observations. 
We obtained two spectra of the star, 35\,d apart in time, and at
first sight they appear to be two different stars (see upper two spectra in
Figs.\,\ref{fig:HD96237a} and \ref{fig:HD96237b}). This is, however, not the case.
Radial velocity measurements give the same velocity, 1.4\,\kms, within 340\,\ms.
 As shown in the figure, the overall shape of 
\halpha\ is similar, considering the $\sim 4$ per cent uncertainty in the
normalisation of this region of the UVES spectrum (two orders merge here), 
and a possibly very inhomogeneous surface distribution of elements.
An 8000\,K model spectrum fits the \halpha\ wings best in the 
2007 February spectrum, and  
a slightly cooler ($\sim200$\,K) model spectrum of about 7800\,K fits the 
\halpha\ wings better in the 2007 March  spectrum. Both temperatures agree 
well with the late-A spectral type indicated by the shape of the \ion{Ca}{ii}\,K-
line 
profile.  The stable radial velocity and single \halpha\ line (also other 
hydrogen lines and calcium lines in the FEROS spectrum are single) give 
no evidence of a secondary spectrum. The 2007 March spectrum was recorded 
during evening twilight at an air mass of 1.8,  but line contamination from 
the solar spectrum is excluded as the reason for the difference from the 
2007 February  spectrum. 

\begin{figure}
 \vspace{-1pt} 
 \hspace{-16pt}
\includegraphics[width=0.51\textwidth, angle=0]{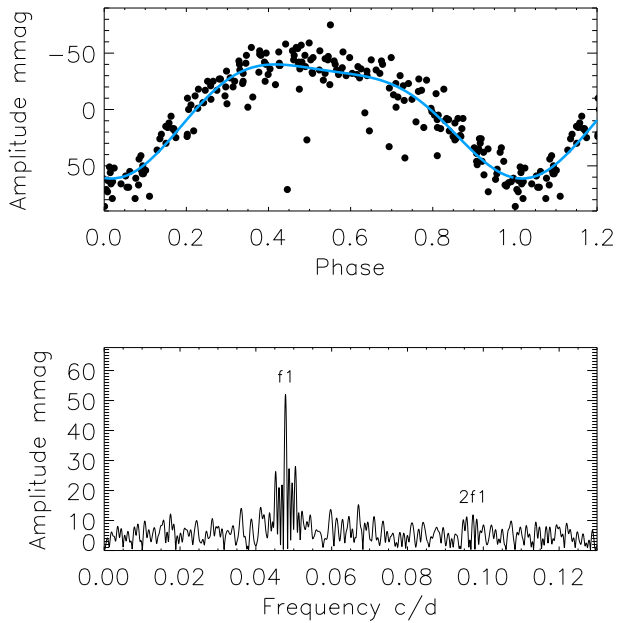}
 \caption{\label{fig:HD96237}
 ASAS light curve of HD\,96237 folded with the period of 20.9\,d.
 Note the light curve's departures from a purely sinusoidal
 variation, which is reproduced well by including the first
 harmonic in the fit (superposed). Both frequencies are indicated
 in the amplitude spectrum in the bottom panel. 
 }
\end{figure}

The star is a known photometric variable listed as $\alpha^2$\,CVn in the 
General Catalog of Variable Stars (GCVS4.2, \citealt{samusetal04}), and as a 
semi-detached eclipsing binary in the ASAS catalogue. Period analysis
of the ASAS light curve confirms a variability (Fig.\,\ref{fig:HD96237}), 
which we interpret as a single-wave $\alpha^2$\,CVn periodicity of 20.91\,d with 
an amplitude of 49\,mmag. As demonstrated in Fig.\,\ref{fig:HD96237}, the fit
is improved by including the first harmonic at 0.0956\,cd$^{-1}$
which is detected at the 4.4\,$\sigma$ significance level.
The {\it Hipparcos} light curve confirms the 20.91\,d period. 
Assuming this to be the rotation period (we cannot rule out
a double-wave light curve with a 41.83\,d rotation period)
it agrees well with the short timescale between the two epochs of our recorded 
spectra, during which the abundances have changed.
The spectrum variability is exceptional among known Ap stars, and the very
peculiar line strengths in the 2007 February  spectrum resemble those of the
most peculiar Ap star known, HD\,101065 
(Figs.\,\ref{fig:HD96237a} and \ref{fig:HD96237b}). That 
latter star has, however, a rotation velocity close to zero, which makes HD\,96237
of particular interest for studying the chemical surface distribution in 2-D. 

HD\,96237 now joins the small group of extreme Lanthanide Ap stars 
that 
includes HD\,51418 \citep{jonesetal74} and the more famous HR\,465 (HD\,9996; 
\citealt{prestonetal70}), as well as HD\,101065. These other stars have
comparable magnetic field strengths:
HD\,101065 has partially resolved Zeeman 
components giving a magnetic field modulus of $\left<B\right>=2.30\pm0.35$\,kG
\citep{cowleyetal00} in agreement with $\langle B \rangle = 2.3$\,kG measured 
by us using UVES spectra and
based on a partially resolved Zeeman pattern in the \ion{Gd}{ii}\,5749\,\AA\ line;
HD\,51418 has a mean longitudinal magnetic field which varies from 
$\langle B_{\rm z}\rangle=-200$  to 750\,G 
\citep{jonesetal74}; and 
HR\,465 has $\left<B\right>=4.83\pm0.39$\,kG \citep{mathysetal97}.
HD\,51418 has a rotation period of 5.4379\,d, typical of the Ap stars, 
but HR\,465 is known for 
its long rotation period of $22-24$\,y and extreme spectral variations -- from a 
typical Ap star to an extreme lanthanide star at rare earth element maximum. In 
the 
sense of extreme rotational spectral variations, then HD\,96237 is comparable to 
HR\,465, but has the tremendous advantage of a shorter 20.9-d rotation period, 
making its study in detail much more practical.  

Figure\,\ref{fig:fe6149a} shows a triple line structure at the location of
$\lambda$6149\,\AA\  that is {\em not} caused by magnetic splitting. (The 
$\lambda$6149\,\AA\
splitting would here have corresponded to a 6\,kG field, which 
is inconsistent with other lines.) Comparison
with the 2007 March  spectrum (Fig.\,\ref{fig:fe6149b}) shows that the blue-most
component has disappeared while the two components centred around the 
$\lambda$6149\,\AA\ 
line are replaced by a single, deeper, rather broad line. Our abundance estimates 
(Table\,\ref{tab:abundance}) show a considerable increased iron abundance in
the latest spectrum, while abundances of Nd and Pr 
have decreased on average more than 1.1\,dex (Eu increased the same amount).
The splitting of $\lambda$6149\,\AA\ in the 2007 February  
spectrum is therefore most 
probably due to strong unidentified lines of, e.g., rare earths and a relatively
low iron abundance.  HD\,101065 (Fig.\,\ref{fig:HD96237a} and \ref{fig:HD96237b}) 
is a known case with precisely that condition. 
Abundances of Cr, Pr, Nd and Eu are greater than solar in both our HD\,96237 spectra, 
while
Fe is solar or near-solar. Ba $\lambda$6141\,\AA\  is present in both spectra,
and enhanced strongly in the most recent one.

\begin{figure}
 \vspace{1pt} 
 \hspace{1pt}
\includegraphics[height=0.41\textheight,width=0.44\textwidth, angle=0]{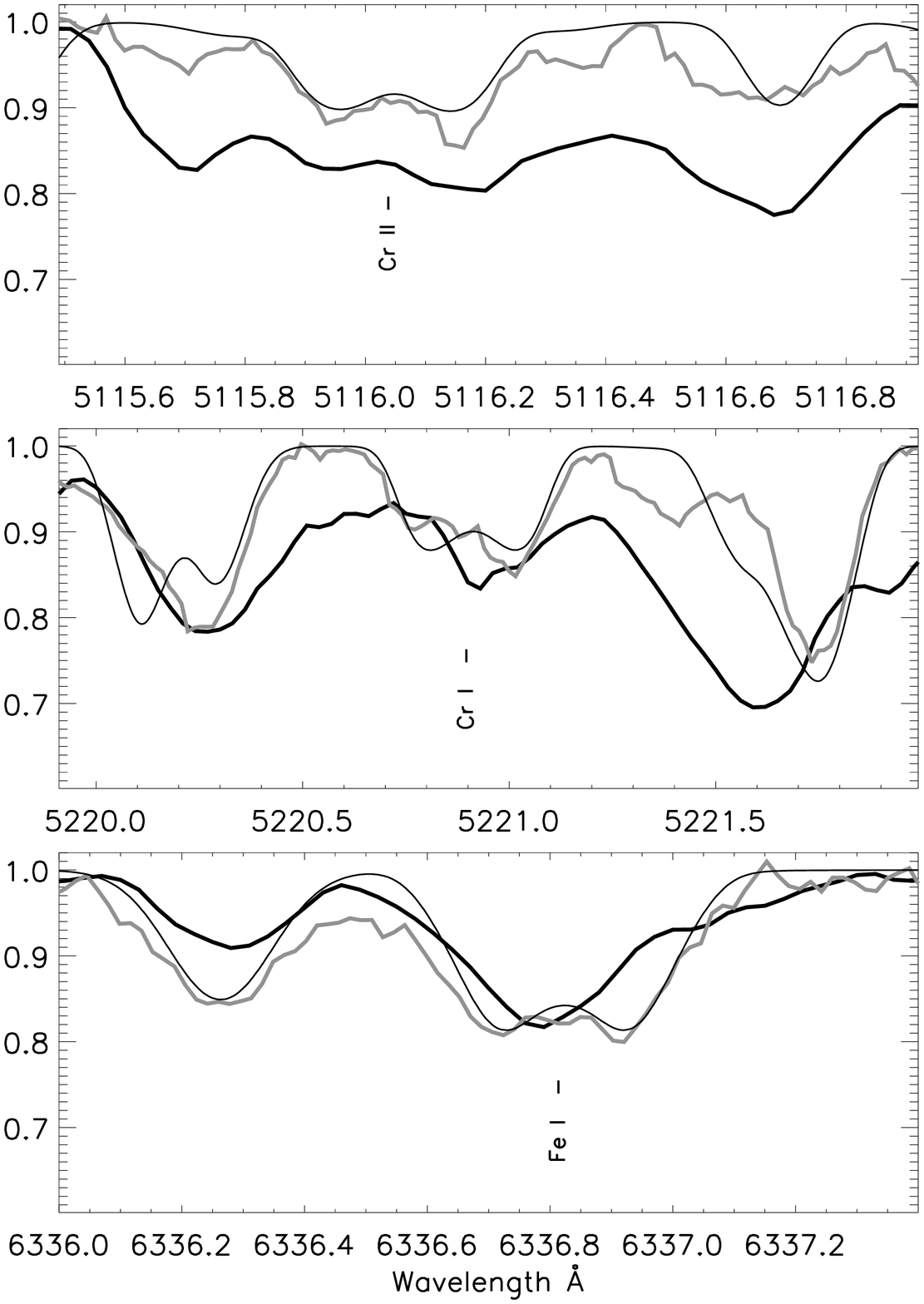}
 \caption{\label{fig:HD96237mag}
Magnetic field of HD\,96237. The magnetically sensitive lines
\ion{Cr}{i}\,5116.049\,\AA, \ion{Cr}{ii}\,5220.912\,\AA\  and
\ion{Fe}{i}\,6336.824\,\AA\  (top to bottom panels) of the 2007 February 
(thick black line), 2007 March  (grey line) and a model spectrum superposed for
a 3.2\,kG field (3\,kG radial and 1\,kG meridional field). Land\'e
factors of the three lines are $g_{\rm eff}=2.92$ ($\lambda$\,5116.049), 
3.00 ($\lambda$\,5220.912) and 2.00 ($\lambda$\,6336.824). Direct
measurements of the splitting of the 2007 March  spectrum gives the
corresponding magnetic field strengths $\left<B\right>=2.5$, 2.8 and 2.4\,kG.
The model is rotationally broadened to \vsini\,=\,4.5\,\kms. Ordinate is 
normalized intensity.
 }
\end{figure}

HD\,96237 does in fact have a $\left<B\right>\sim3$\,kG magnetic field which is 
responsible for the
magnetic broadening of  the iron $\lambda$6149\,\AA\  line in 
Fig.\,\ref{fig:fe6149b}.
The high-resolution 2007 March  spectrum reveals lines with partial Zeeman 
splitting and direct measurements of the splitting of \ion{Cr}{i}\,5116.049\,\AA, 
\ion{Cr}{ii}\,5220.912\,\AA\  and \ion{Fe}{i}\,6336.82\,\AA\  indicates a field of
$\left<B\right>=2.87\pm0.30$\,kG, while a two-Gaussian fit to the 
 $\lambda$6149\,\AA\ line indicates $\left<B\right><3.27$\,kG. 
A {\small SYNTHMAG} model with a 3.16\,kG field ($B_{\rm r}=3.0$\,kG  
and $B_{\rm m}=1.0$\,kG) is compared to these lines in
Fig.\,\ref{fig:HD96237mag}. This quantity should be
comparable to the mean magnetic field modulus, but in general not
exactly equal to it as it is model dependent while $\left<B\right>$
is not. The same figure shows the 2007 February spectrum superposed
the latest spectrum and although the Cr lines are not directly
incompatible with the field strength, iron lines are useless in
this spectrum due to this element's relatively low abundance.
A single polarimetric measurement of HD\,96237 was kindly obtained 
on 2008 January 23 (JD2454488.534) by Dmitri Kudryavtsev 
and Iosif Romanyuk using the Russian 6-m telescope at the SAO/RAS.
(see \citealt{kudryavtsevetal06}
for a description of data reduction procedures and instrumentation.)
It confirms the magnetic field with a longitudinal field 
 of $\langle B_{\rm z}\rangle$$=-720\pm70$\,kG.
We are currently obtaining additional spectra of this important 
object.

\subsubsection{HD\,110274} 

\begin{figure}
\includegraphics[width=0.48\textwidth, angle=0]{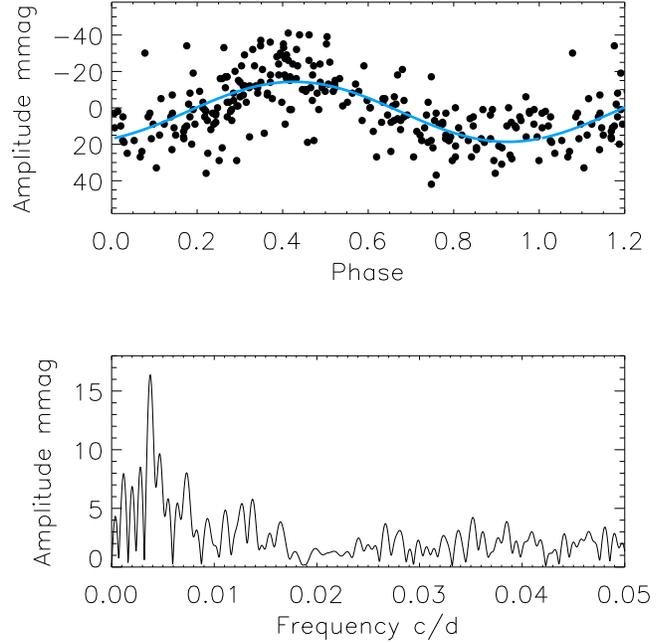}
 \caption{\label{fig:HD110274}
 ASAS light curve of HD\,110274 folded with the period of 265.3\,d.
 Below is the corresponding amplitude spectrum.
 }
\end{figure}

Because of the small rotational broadening of HD\,110274, \vsini\,=\,1\,\kms, 
$\lambda$6149\,\AA\ is magnetically resolved, in spite of a relatively 
weak field.
After obtaining two spectra of the star with FEROS, we confirmed the discovery 
with a single UVES spectrum. The sharp-lined spectra have identical 
radial velocities for the FEROS spectra, ${\rm RV}=-6.04$\,\kms, while 30\,d
later, the UVES spectrum appears slightly altered, ${\rm RV}=-6.59$\,\kms. 
Considering that the wavelength calibrations were only checked with telluric 
lines, the common errors of the measurements overlap, and the star may be
radial velocity  stable. With the $\lambda$6149\,\AA\ line we measure a magnetic 
field strength of 3.80, 3.80 and 4.45\,kG in chronological sequence 
($\left<B\right>=4.02\pm0.38$\,kG, combined). This increase appears 
significant: with a total of 3 Fe, Cr lines (9 for the UVES
spectrum, including Eu), we get correspondingly the  measurements
$\left<B\right>=3.55\pm0.22$, $3.73\pm0.57$ and $4.55\pm0.50$\,kG.
The ASAS light curve shows a 265.3\,d periodicity with 16\,mmag
amplitude (11.2\,$\sigma$, Fig\,\ref{fig:HD110274}). {\it Hipparcos}
photometry confirms this periodicity (249.5\,d, 5.0\,$\sigma$).
The spectra show many continuum windows and lines of Ba, Ca and \ion{Si}{i} are
present, though weak. Cr, Nd, Pr, Eu are greater than solar, Fe is 
near-solar.

\subsubsection{HD\,117290} 

For a clearly split  $\lambda$6149\,\AA\ line, we measure a constant magnetic 
field 
strength of 
6.40, 6.38 and 6.36\,kG, chronologically for three spectra
($\left<B\right>=6.38\pm0.02$\,kG, combined). 
The last measurement is separated by 30\,d from the first two.
Adding further $4-6$ Fe, Cr, Eu, Nd and Ce lines to
the analysis we get, within the errors, an unchanged field of
$6.27\pm0.82$\,kG for all spectra combined. The radial velocity of HD\,117290 is 
also 
constant (${\rm RV}=-28.9\pm0.2$\,\kms) for the two spectra  obtained on 
HJD\,2454139 and 2454171. 

\begin{figure}
 \hspace{-1pt}
\includegraphics[width=0.46\textwidth, angle=0]{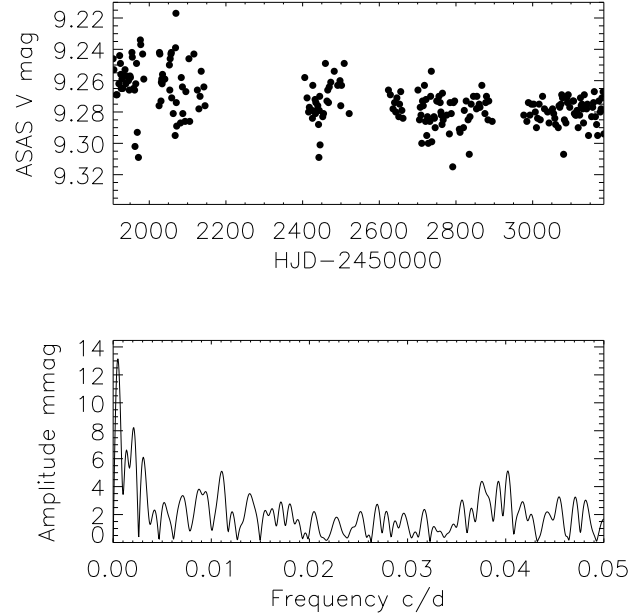}
 \caption{\label{fig:HD117290}
 The 3.8\,yr long ASAS light curve of HD\,117290.
 A 3-$\sigma$ filter was applied to reject
 outliers from the light curve.  Below is the corresponding amplitude spectrum.
 }
\end{figure}

The  3-$\sigma$ filtered ASAS photometry shows a 30\,mmag decrease in brightness
over 3.8\,y (Fig.\,\ref{fig:HD117290}). The photometric scatter is considerable in the
first half of the light curve and with no {\it Hipparcos} data for this star,
more data for diaphragms smaller than those of ASAS are needed to confirm the trend. 
If this is associated with the stellar rotation, the shortest plausible period 
for a sinusoidal variation is 5.7\,yr. Continuum windows are plentiful.  Ba, Ca 
and Si are all overabundant, as are Cr, Nd, Pr, Eu, while Fe is 
solar.

\subsubsection{HD\,121661}

\begin{figure}
 \vspace{-3pt} 
 \hspace{-13pt}
\includegraphics[width=0.49\textwidth, angle=0]{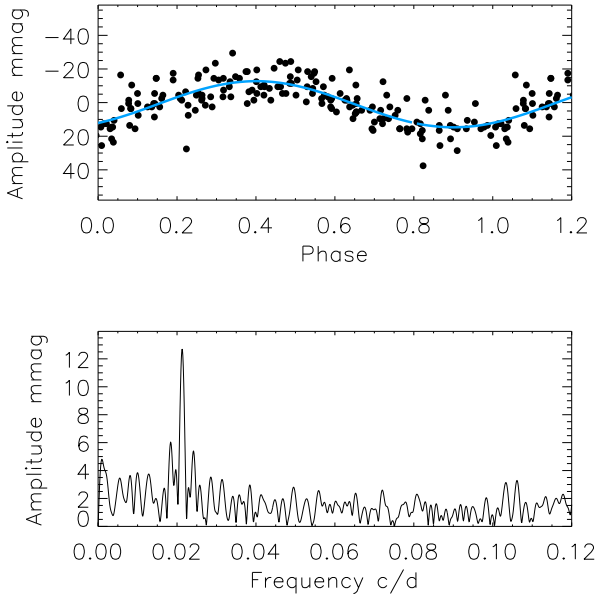}
 \caption{\label{fig:121661}
 ASAS light curve of HD\,121661 folded with a 47.0\,d period.
 Below: the corresponding amplitude spectrum.
 }
\end{figure}

This star was observed twice, separated by 30\,d.
From  $\lambda$6149\,\AA, we measure in sequence a magnetic
field strength of $\left<B\right>=5.41$ and 6.92\,kG 
(or $6.16\pm1.07$\,kG, combined). 
Compared to the small error, $\sigma(\left<B\right>)=10$\,G, for the 
weaker magnetic field of HD\,52847 measured from a relatively similarly shaped
 $\lambda$6149\,\AA\ line and comparable span in time, the 1.51\,kG change for 
HD\,121661 is significant. From Figs.\,\ref{fig:fe6149a} and \ref{fig:fe6149b}
one may also notice a considerable change in the Fe line's shape.
With 7 
Fe and Cr lines (8 for the UVES spectrum, including Eu), we get   
$\left<B\right>=5.40\pm0.47$ (2007 February) and 
$\left<B\right>=6.01\pm0.82$\,kG (2007 March).
\changea{
The radial velocity decreases marginally 
by about 2\,$\sigma$: $1.33\pm0.66$\,\kms\ (from ${\rm 
RV}=7.35$\,\kms\ to ${\rm RV}=6.02$\,\kms, in sequence), but}
this needs to be confirmed before concluding the star is a radial velocity 
variable.
The ASAS light curve in Fig.\,\ref{fig:121661} shows a 47.0\,d periodicity 
(14\,mmag amplitude, 13.7\,$\sigma$) which is interpreted as the rotation period.
Ba (including the $\lambda$6141\,\AA\ line), \ion{Si}{i} and Ca are absent or weak
 and many continuum regions are visible. Cr, Nd, Pr, Eu have  greater than solar
 abundances while Fe is near-solar. 

\subsubsection{HD\,135728AB} 
\begin{figure}
 \hspace{-2pt}
\includegraphics[height=0.465\textwidth, angle=90]{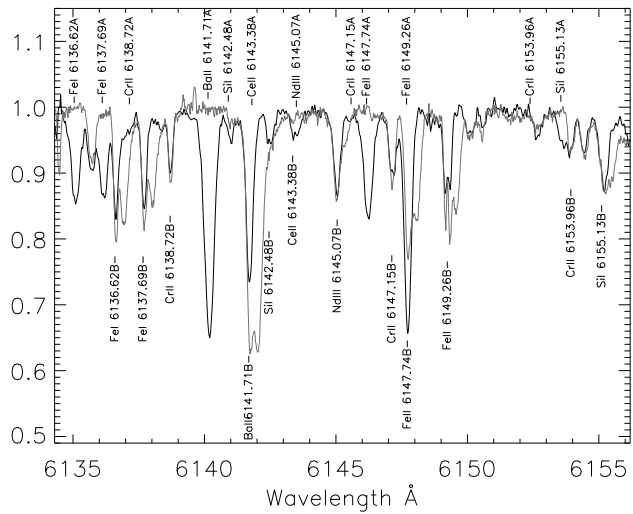}
 \caption{\label{fig:HD135728}
 A region of our two 
 spectra of the SB2 binary HD\,135728AB, with the magnetic component 
 shifted to rest wavelengths. Ordinate is normalized intensity. In the 2007 February
 spectrum (black), the primary star's spectrum is shifted to shorter
 wavelengths than the secondary and to longer wavelengths in the 2007 March 
 spectrum (grey). Locations of lines are indicated for
 both components in the 2007 February spectrum (black). Those referring
 to the broad-lined primary HD\,135728A are indicated above the 
 continuum (identifications ending on `A'), and below the continuum for
 the secondary HD\,135728B  (identifications ending on `B').
 Note, e.g., the stronger Ba line of HD\,135728A. Nd is strongest 
 in HD\,135728B and this star's magnetically resolved Fe line at 6149\,\AA\ is 
 obvious in both recorded spectra.
 }
\end{figure}

As a new SB2 binary star discovery, this star is of particular interest by 
having a magnetic component. The magnetic component has the 
next-to-weakest field in our FEROS study and was re-observed 
32\,d later with the  high resolution of UVES. At that moment,
the system was near orbital quadrature so the lines of the binary components 
are mostly merged, but the splitting of $\lambda$6149\,\AA\ is
evident in both spectra (Fig.\,\ref{fig:HD135728}). 
We measure in sequence the following radial velocities in the two spectra: 
$-60.3\pm0.7$ and $-17.2\pm0.4$\,\kms\ 
for the primary, and $+14.3\pm0.5$ and $-32.7\pm0.1$\,\kms\ for the secondary. 
With only two spectra, interpretation of dimensions of HD\,135728 component's
is speculative. Nevertheless, to get a first impression of the system, 
we 
assume a circular orbit and not too different temperatures of the two stars. For
this working hypothesis we here label the broader lined component 
(see Fig.\,\ref{fig:HD135728}) the primary star,
 HD\,135728A, as it appears to be more luminous and massive (see below),
and the sharp-lined, magnetic component for the secondary star HD\,135728B.
The maximum observed radial velocity separation of the two 
stars, 74.6\,\kms, indicates a relatively close orbit.
The component spectra have
exchanged relative locations in our two spectra, and the 
estimated systemic
velocity is $V_\gamma \sim  -24.6$\,\kms.
The resulting observed RV amplitudes $a_{\rm B}$ and
$a_{\rm A}$ thus constrain the orbital velocity amplitudes to
$K_{\rm A}\ge35.7$\,\kms\ and  $K_{\rm B}\ge38.9$\,\kms, leading to the mass ratio 
$q=M_{\rm A}/M_{\rm B}=K_{\rm B}/K_{\rm A} \sim a_{\rm B}/a_{\rm A}=1.09$.
The maximum period of the orbit can then be derived from:
\begin{equation}
M_{\rm A} + M_{\rm B} =
\frac{(1-e^2)^{3/2}}{2 \pi G \, \sin^3 i} (K_{\rm A} + K_{\rm B})^3 P \; .
\end{equation}
By assuming: no eccentricity $e=0$; the sum of the radial velocity amplitudes
$K_A+K_B > 74.6$\,\kms; a mass of HD\,135728A of
$M_A\sim2.2\,\mathrm{M}_{\odot}$ (corresponding to the
spectroscopic and photometric temperatures); and inclination
$i<85\;\mathrm{deg}$ (no eclipses are seen),
then $P<96$\,d. The true period can be an order of magnitude less,
depending on the maximum radial velocity amplitudes. 

At large separation, HD\,135728B's  $\lambda$6149\,\AA\ line indicates a 
field of $\left<B\right>=3.89\pm0.11$\,kG, and together with 
\ion{Cr}{i}\,5247.57\,\AA\ and \ion{Fe}{i}\,6336.82\,\AA\ a field of
$\left<B\right>=3.54\pm0.49$\,kG is found. 
For the high-resolution 2007 March spectrum, the same 3 lines 
(here partly blended) provide 
$\left<B\right>=3.43\pm0.52$\,kG, or 
$\left<B\right>=3.37\pm0.28$\,kG for $\lambda$6149\,\AA\ alone.
This suggests a constant field strength of 
$\left<B\right>=3.63\pm0.30$\,kG based on the $\lambda$6149\,\AA\ line.
We cannot exclude a magnetic
field of up to $\sim 2.5$\,kG for component A, which could be evaluated
from a detailed analysis of magnetic broadening or moments of line profiles,
such as described in \citet{mathysetal06} and \citet{freyhammeretal08}. 

The two component spectra are similar in the covered wavelength range
(e.g. \ion{Si}{ii}\,6347 and 6371\,\AA, \ion{Mg}{ii}\,4481\,\AA, 
\ion{Ca}{ii}\,8498\,\AA), though with a different light ratio.
The temperatures of the two stars therefore appear to be similar,
Which, e.g., for the \ion{Ca}{ii}\,K region corresponds to
an early-F type star. The photometric temperature for the combined light
indicates a much higher temperature ($\teff=8060$\,K).
The ASAS $V$ photometry shows no long-period variability above 6\,mmag. 
We estimate from the double \halpha\ line that HD\,135728B only contributes to the 
total light with $37\pm3$ per cent. 
This estimate was made by comparing a composite of two identical spectra,
by applying only a wavelength shift and different light factors. 
Direct fitting of Gaussians to the blended \halpha\ 
cores gives a similar light factor of the secondary star (35 per cent).
The corresponding
light ratio is $\sim1.70$, which for the luminosity of the system in
Table\,\ref{tab:targets} leads to estimated individual luminosities of 
$\log L_{\rm A}\sim1.40$ and $\log L_{\rm B}\sim1.17$\,L$_\odot$, in reasonable 
agreement with the preliminary mass ratio. 

Both components have greater than solar abundances of Cr, Nd, Pr and Eu,
but considerably
stronger rare earth elements in the secondary, magnetic, component (HD\,135728B). 
Fe has near-solar abundance in the A component, and greater than solar
 abundance in the
other star. We emphasise that the estimates in Table\,\ref{tab:abundance} are
based on the much simplified assumption of same luminosities of the two stars. 
Ba $\lambda$6141\,\AA\ is strong in both stars but considerably stronger in 
HD\,135728A.  HD\,135728B has a weaker \halpha\ line,    is 
more sharp lined (\vsini\,=\,2\,\kms) and its $\lambda$6149\,\AA\ line is 
magnetically resolved.
Ca is of similar strength in both stars, about  $-5.34$ dex. Also Sc is clearly
present in HD\,135728B, while only \ion{Sc}{ii}\,5657\,\AA\ was identified in HD\,135728A.
The abundances of Sc also seem similar, which lends some support to HD\,135728A being Ap.
Neither of the stars have Nd or Pr ionization disequilibrium anomalies.

\subsubsection{HD\,143487}
With a rather peculiar spectrum, this star is a particularly
good roAp candidate.
Comparing the 2007 February spectrum with the 
2007 March (UVES) spectra, we find no change in the absolute radial velocity
of the star (${\rm RV}=-26.0$\,\kms). The Zeeman splitting of $\lambda$6149\,\AA\
indicates a constant magnetic field of $\left<B\right>=4.23\pm0.07$\,kG for all 
spectra. Including
two more Cr lines, the composite UVES spectrum (Table\,\ref{tab:obslog}) shows a
field of  $\left<B\right>=4.30\pm0.16$\,kG.
No long-period variability above 6\,mmag is detected in the photometry.

The spectrum of HD\,143487 is highly peculiar and, e.g., Pr, Nd
are among the strongest in this study. Also Cr and Eu abundances are
greater than solar, while that of Fe is solar. Ba\,$\lambda$6141\,\AA\  is very 
shallow ($<5$ per cent below continuum). There is a clear
ionization disequilibrium anomaly of Nd and Pr with
$\Delta[{\rm Pr}]_{\rm III-II}=1.50$ and $\Delta[{\rm Nd}]_{\rm
III-II}=1.36$,
in excellent agreement with empirical findings
for roAp stars \citep{ryabchikovaetal04}. With those anomalies and its 
relatively low temperature
$T_{\rm  eff} = 6930$\,K, HD\,143487 is a prime roAp star candidate.

\begin{figure}
\hspace{-0.3cm}
\includegraphics[height=0.35\textheight,width=0.48\textwidth, angle=0]{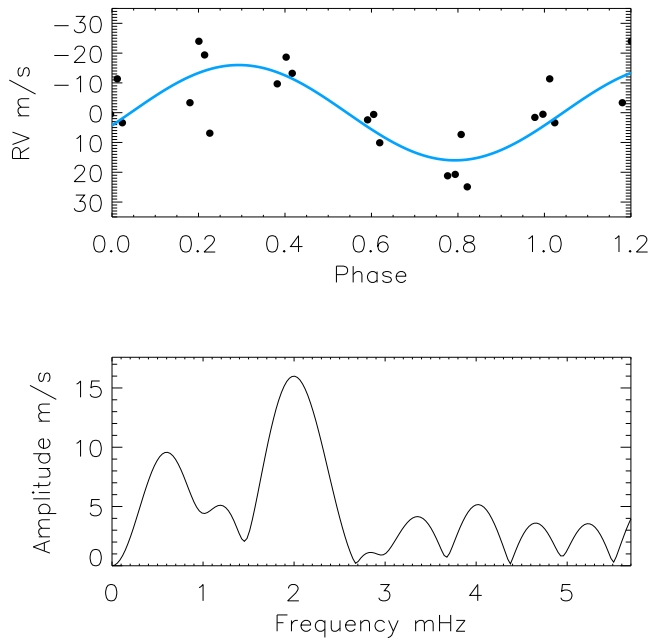}
 \caption{\label{fig:HD143487} An initial probe for rapid pulsations
in the radial velocities of 18 HD\,143487 spectra. The top panel shows our
32\,min radial velocity curve folded with a 2\,mHz candidate period. 
The velocities are from cross-correlation of all
spectra with an average spectrum. 
The bottom panel shows the corresponding amplitude spectrum.
The data are de-trended for a linear drift. While the peak at 2\,mHz is plausibly 
real, it needs confirmation.}
\end{figure}

We attempted detection of pulsations with UVES by obtaining 18
spectra in 32\,min. Using the procedures described in 
\citet{freyhammeretal08} the spectra were then searched for 
rapid pulsations, but due to the short run the results were
inconclusive. An intriguing result, however, was that cross-correlation
of the spectra with the average spectrum in the 
$5150-5860$\,\AA\ region indicates (Fig.\,\ref{fig:HD143487})
a 2\,mHz (8 minute) period at a 4.6-$\sigma$ level. This, however,
could not be confirmed by detection of radial velocity variations in 
individual lines or groups of combined lines. Nevertheless, as the case of 
$\beta$\,CrB shows (\citealt{hatzesetal04}; \citealt{kurtzetal07}),
it is possible to detect roAp pulsation 
using large chunks of the spectrum when individual lines may not have sufficient 
$S/N$ to show a clear signal. Fig.\,\ref{fig:HD143487} makes a case for HD\,143487 
being a new roAp star, but we conservatively await confirmation of this. More data 
are now being collected for this purpose.

The spectral classification from the Michigan Catalogue with a note is {\it *APEC. 
Overlapped; many lines; probably Sr, Eu, Cr, but many others also.}

\section{Discussion and conclusions}

As part of a systematic search for new roAp candidates using 
FEROS on the ESO 2.2-m telescope, we
have discovered 17 new magnetic stars with magnetically resolved lines
and with these lines measured their mean magnetic field
moduli directly.  For 11 stars, spectra were obtained about 30\,d
later at higher resolution with UVES on the VLT. These were used 
to confirm the discoveries
and check the stability of the measured magnetic field strengths 
and radial velocities. 

A new double-lined spectroscopic binary, HD\,135728AB, 
was discovered with two similar 
components, for one of which, the more slowly rotating component, 
a magnetic field was detected. It is possible that the primary (the faster
rotating and more massive star, as deemed from its significantly 
smaller radial velocity difference between our two spectra) 
is either an Am or 
Ap star -- i.e. either magnetic or non-magnetic; as yet we cannot 
tell.  It is 
likely that the secondary is a spectrum variable (the primary may be also) so that 
the rotation periods of the stars can be determined independently and compared to 
the orbital period, thus showing whether either or both of the stars are 
synchronously rotating. 
Since both stars show overabundances typical of Ap and Am stars, a 
first guess might be that one is Am (non-magnetic) and the other Ap (magnetic). If 
that is so, how can two, rather similar stars in a close binary with both in the Ap-Am 
domain end up with one strongly magnetic and the other not? 
Or, on the other hand, if the 
primary {\it is} magnetic, but has a weaker field, then the question is similar, 
but not so extreme: how do the two components end up with different field 
strengths.
There are Am-Am SB2 systems known -- WW\,Aurigae 
is a famous eclipsing example where the stars are very similar in mass, but not 
identical in abundances. 
SB2 systems with a magnetic Ap star are very rare and
HD\,135728AB may be particularly promising 
for illuminating the magnetic field origin question.
Examples of other such cases are: 
HD\,59435 \citep{wadeetal99}, HD\,55719 \citep{bonsack76} and
HD\,98088 \citep{abtetal68,hensberge74,bychkovetal05}.  
These binaries have the following characteristics:
magnetic periods of 5.8 -- 1360\,d, orbital periods of 5.9 -- 1386.1\,d,
magnetic fields of 1.45 -- 8\,kG and companion-to-Ap star mass ratios of 0.75 -- 1.33.
In a recent study of the very young binary HD\,72106, \citet{folsometal07} 
established a dipole field of 1.3\,kG for the primary star which rotates fast
enough to permit these authors to produce 2-D surface abundance maps.

It is possible that two other systems in our sample
are binary: HD\,55540
had a significant radial velocity  change, while HD\,121661
needs confirmation of a marginally significant change. Neither of these systems, 
if confirmed, are of the importance of HD\,135728AB and 
further observations are planned to determine the orbital period 
and use spectral disentanglement to study the abundances of 
both its components in detail.

The most important discovery, HD\,75049, has the
second-largest known magnetic field of any Ap star and was found to be highly
variable in magnetic field strength and fine-structure over a time scale
of 30\,d. The extremely strong magnetic field may not only rival, but even 
surpass the strength of Babcock's star (HD\,215441), 34.4\,kG. 
Follow-up is in progress of this very interesting object.

HD\,96237  was shown to exhibit extreme abundance variations, possibly
related to a photometric variability of 22\,d that may be the rotation
period. Extreme abundance variations with stellar rotation are known from 
more slowly rotating stars, such as HR\,465 (HD\,9996; \citealt{prestonetal70}) 
for which Eu, Cr, Ca, Sr vary up to a factor of 3 in line strength
(Cr and Eu in antiphase). A magnetic field was detected and measured. 
We compared the spectra of HD\,96237 with those of
the arguably most peculiar Ap star, HD\,101065, and demonstrated
comparable levels of peculiar abundances. HD\,96237 is remarkable by
also exhibiting fast abundance variations. Follow-up studies 
are currently in progress of this important star.

From light curves in the ASAS database, the stars HD\,75049, HD\,88701,
HD\,96237, HD\,110274, HD\,121661, HD\,117290 and HD\,92499 were shown to be
$\alpha^2$\,CVn variables. Two periods were detected for 
HD\,75049, while HD\,88701 exhibits a clear double wave.
HD\,117290 and HD\,92499 show, as the only cases, variability longer
than the time span covered by the photometry (3--5 years).
There are some implications of the rotation periods 
found from the ASAS data and the measured rotational 
velocities, since these 
constrain the stellar radii and thus luminosities. An example 
we discussed is HD\,88701 for which the rotation period and \vsini\ implied
an unexpectedly large radius. However, uncertainties in \vsini\ measurements, 
sensitive
to line-blending and magnetic broadening and the possibility of an ASAS
period only being half of the rotation period (in case of double-wave
light curves) leaves some uncertainties. Furthermore,
astrometric luminosities suggest that at least half the stars are 
near the terminal end of the main-sequence, so that larger radii are to be 
expected in comparison with younger Ap stars. 

Included among our original sample 140 cool Ap stars is HD\,92499, a known
magnetic star with Zeeman splitting \citep{hubrigetal07}. Our
new spectra showed a constant magnetic field modulus and radial velocity of
the star. As our target selection is relatively unbiased among more than
500 cool Ap stars, the fraction of such stars with magnetically
resolved lines appears to be 13 per cent based on the first part of
our Ap star survey.

Abundance estimates were used to identify Nd and Pr ionization 
disequilibrium anomalies in abundances of ions in the two first 
ionized states. HD\,44226, HD\,96237 and HD\,143487 showed
significant abundance anomalies ($\sim1$\,dex) and are in addition
to HD\,92499  excellent roAp candidates.
With 32\,min time-series spectroscopy of HD\,143487, we demonstrated
a low-amplitude candidate period of 2\,mHz that, however, 
could not be confirmed by individual lines. 

We emphasise that the lack of accurate 
{\it Hipparcos} \citep{hip} parallaxes $[\sigma(\pi)/\pi<0.2]$ for the
newly detected magnetic stars presented in this paper means that absolute 
magnitudes are considerably more 
difficult to determine because of the peculiar spectra of these stars.
We note that for a subset of the stars having parallaxes, their 
relatively clustered location in the H-R diagram indicates that many of these are
stars near the end of their main-sequence lifetimes.
This new sample of stars with directly measured magnetic fields will aid 
studies of magnetic field interaction with stellar atmospheres. 
Polarimetric
measurements are needed to establish geometry of the detected fields and 
to determine or confirm the rotation periods of the stars.
We are currently obtaining
high time resolution spectroscopy 
with UVES for 16 of the 18 stars in this study to
search for rapid pulsations. The exceptions are HD\,75049, which is too hot to be 
a roAp star, and HD\,135728AB, which is an SB2 system. 

\section*{Acknowledgments}
We thank the referee, Dr Stefano Bagnulo for a careful reading 
of the paper and many suggestions that helped to improve it.
LMF, DWK and VGE acknowledge support for this work from the Particle
Physics and Astronomy Research Council (PPARC) and from the Science and Technology 
Facilities Council (STFC).  We are grateful for a magnetic measurement of 
HS\,96237,
obtained by by Dmitri Kudryavtsev                             
and Iosif Romanyuk using the Russian 6-m telescope at the Special Astrophysical
Observatory of the Russian Academy of Sciences (SAO/RAS).
LMF received support from
the Danish National Science Research Council's project
`Stellar structure and evolution - new challenges from ground and space 
observations', carried out at Aarhus University and Copenhagen University.
We acknowledge extensive usage of the VALD, VizieR, 
SIMBAD, ADS (NASA) databases.

\bsp

%-------------------
\appendix
\begin{figure*}
\section{Additional figures}
\includegraphics[height=0.91\textheight,width=0.95\textwidth, angle=0]{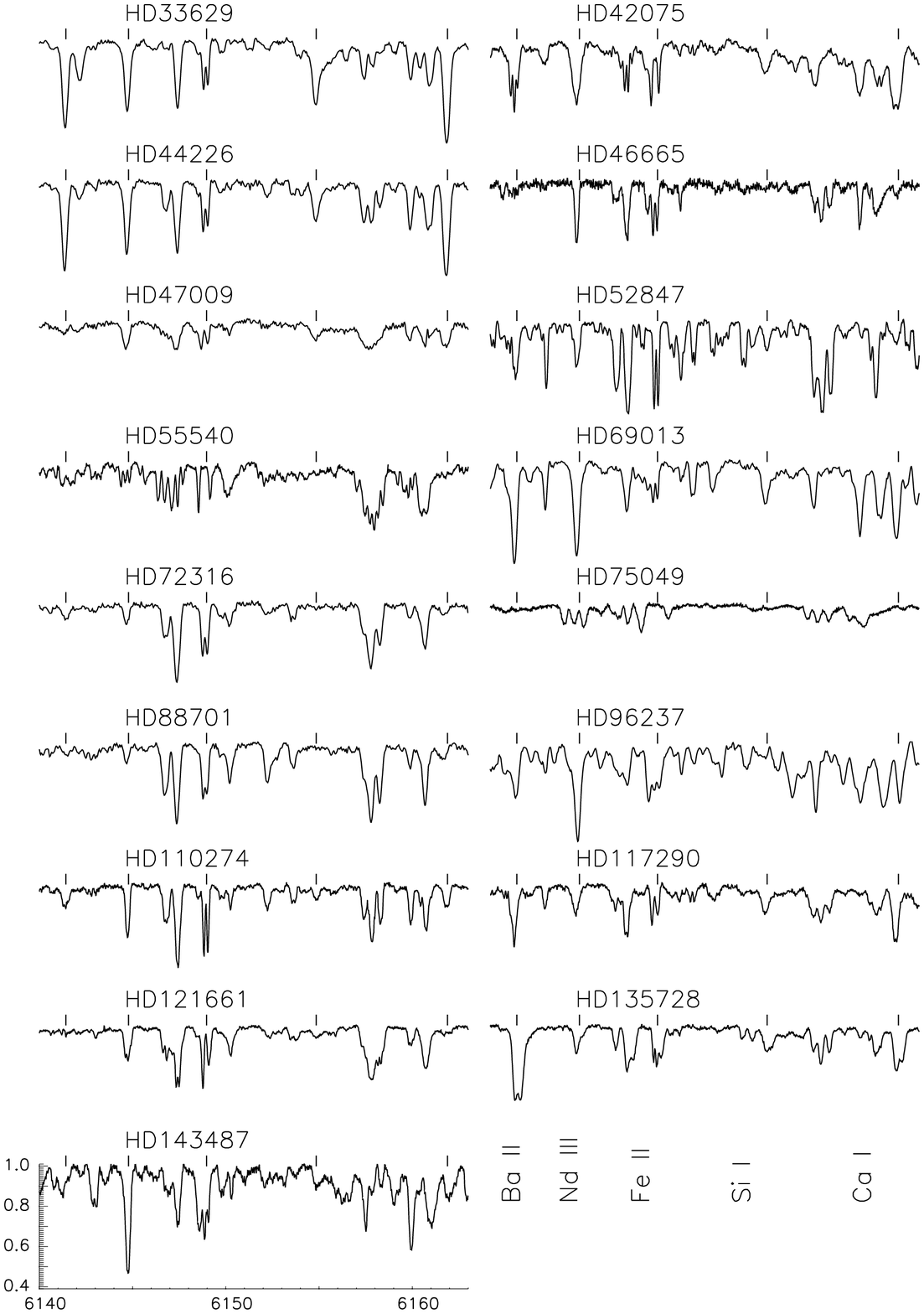}
 \caption{\label{fig:abunall}
 Line strengths for all the new magnetic stars, shown for the same selected 
 region.  All spectra are shifted to the laboratory wavelengths and
 plotted with same scale. Ordinate is normalized intensity. 
 Locations of 5 different lines are indicated.
 }
\end{figure*}
\begin{figure*}
\includegraphics[height=0.91\textheight,width=0.95\textwidth, angle=0]{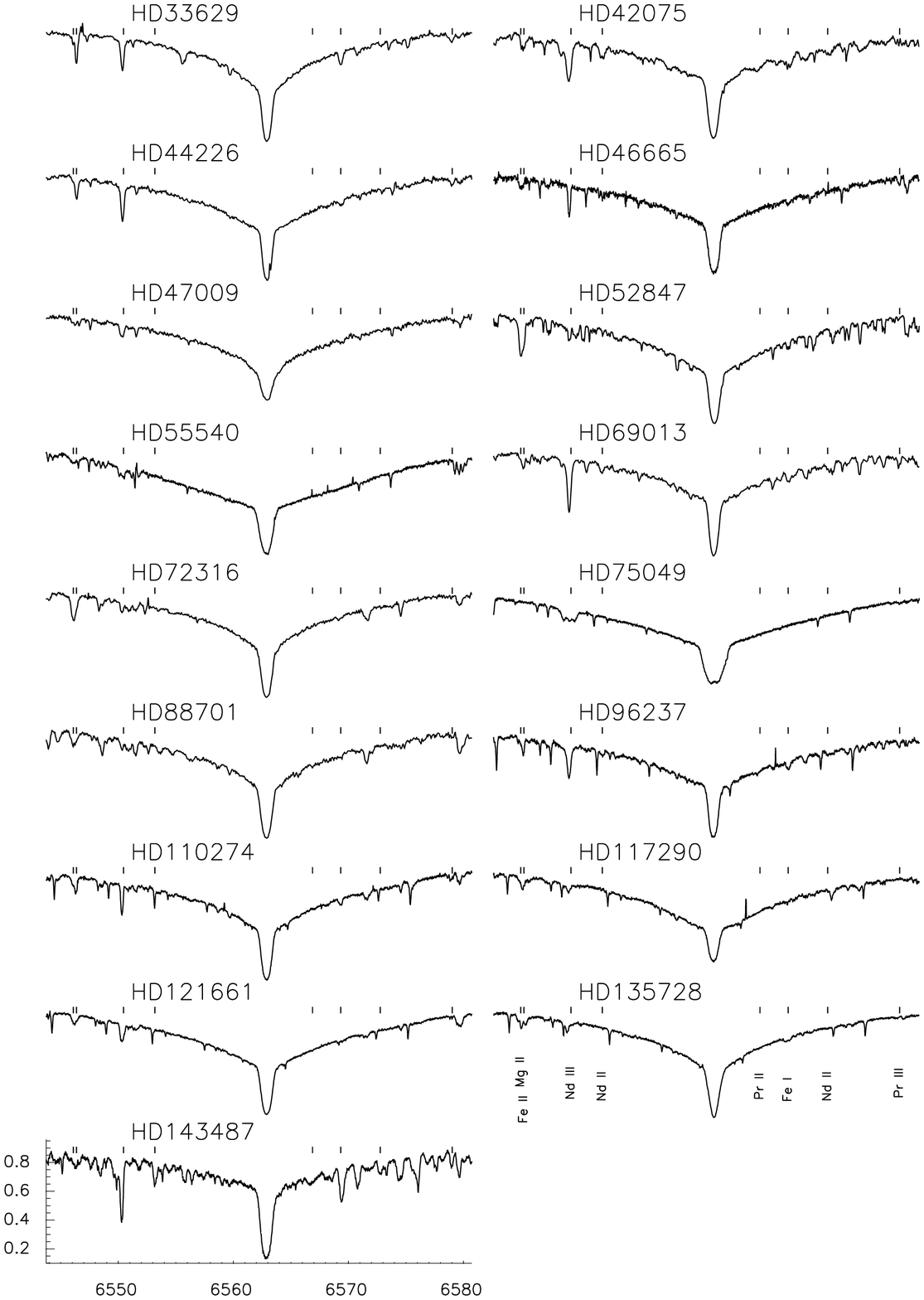}
 \caption{\label{fig:abunhal}
 Same as Fig.\,\ref{fig:abunall}, but for the inner \halpha\  region.
 }
\end{figure*}

\label{lastpage}

\end{document}